\documentclass[aps,prd,groupedaddress,showpacs,showkeys]{revtex4}

\usepackage{amssymb}
\usepackage{amsmath}
\usepackage{amsfonts}
\usepackage{epsfig}

\newcommand{\beaa}{\begin{eqnarray*}} 
\newcommand{\enaa}{\end{eqnarray*}}

\newcommand{\bea}{\begin{eqnarray}}
\newcommand{\ena}{\end{eqnarray}} 

\newcommand{\eq}{\begin{eqnarray}}
\newcommand{\en}{\end{eqnarray}}

\def\arraystretch{1.5}

\begin{document}

\title{Relativistic constituent quark model with infrared confinement}
\author{
Tanja Branz$^{1}$, 
Amand Faessler$^{1}$, 
Thomas Gutsche$^{1}$,  
Mikhail A. Ivanov$^{2}$,  
J\"urgen G. K\"{o}rner$^{3}$,
Valery E. Lyubovitskij$^{1}$ 
\footnote{On leave of absence from the
Department of Physics, Tomsk State University,
634050 Tomsk, Russia} 
\vspace*{1.2\baselineskip}}
  
\affiliation{$^1$ Institut f\"ur Theoretische Physik,
Universit\"at T\"ubingen,\\ 
Kepler Center for Astro and Particle Physics, \\
Auf der Morgenstelle 14, D--72076 T\"ubingen, Germany
\vspace*{1.2\baselineskip} \\
$^2$ Bogoliubov Laboratory of Theoretical Physics,
Joint Institute for Nuclear Research,~141980~Dubna,~Russia 
\vspace*{.4\baselineskip} \\
$^3$ Institut f\"{u}r Physik, Johannes Gutenberg-Universit\"{a}t,
D--55099 Mainz, Germany\\}

\date{\today}

\begin{abstract} 

We refine the relativistic constituent quark model developed
in our previous papers to include the confinement of quarks.
It is done, first, by introducing the scale integration
in the space of $\alpha$-parameters, and, second, by cutting this 
scale integration
on the upper limit which corresponds to an infrared cutoff. 
In this manner one removes all possible thresholds 
presented in the initial quark diagram. 
The cutoff parameter is taken to be the same for all physical processes.
We adjust other model parameters by fitting the calculated
quantities of the basic physical processes to available experimental data.
As an application, we calculate the electromagnetic form factors of the 
pion and the transition form factors of the $\omega$ and $\eta$ 
Dalitz decays.

\end{abstract}



\maketitle

\section{Introduction}
\label{sec:intro}

The analysis of the structure of hadronic matter is one of the key tasks of 
modern particle physics. The generally accepted view is that hadrons are made 
up of the quark and gluon degrees of freedom of Quantum
Chromo Dynamics (QCD) (see, e.g. the monograph~\cite{Thomas:2001kw}). 
An attempt to give a quantitative field theoretic description of hadrons
and their interactions with quarks and gluons  
has to go beyond perturbation theory and necessarily implicates the use of
nonperturbative methods.
Such phenomena as hadronization, i.e. how hadrons are constructed from quarks,
and confinement, the empirical fact that quarks have not been detected 
in isolation, can only be understood via non-perturbative methods. 
Significant progress has been made in constructing various
quark models of hadrons which implement different features of QCD.
For example, potential models provide simple tools that allow one
to describe the hadron spectrum. However, the use of quantum mechanical 
potential models cannot explain the confinement of light quarks 
because quark creation and annihilation effects are essentially 
nonperturbative. In the approaches based on quantum field theory
one can understand quark confinement as the absence
of quark poles and thresholds in Green's functions and matrix elements.  

In the Quark Confinement Model (QCM) Ref.~\cite{Efimov:1993ei}
quark confinement was implemented by assuming that, 
at low energies, the constituent quark interacts with some given vacuum gluon 
configurations. As a result the quark has no fixed pole mass and there are
no quark poles in the Green's functions and matrix elements.    
Models based on results obtained via QCD's Dyson-Schwinger equations (DSEs) 
\cite{Roberts:2000aa} possess 
the feature that quark propagation is described by fully dressed Schwinger 
functions. Dressing of the quarks eliminates the threshold problem and thus
one is effectively working with confined quarks.
Using the framework of quantum field theory a promising approach for the 
description of composite  particles
as bound states of their constituents was suggested in \cite{SWH}.

Over the past years we have developed a relativistic constituent quark model 
and have applied the model to a large number of elementary particle 
processes~\cite{light_mes}-\cite{Bc}. The relativistic constituent quark model
can be viewed as an effective quantum field approach to hadronic interactions 
based on an interaction Lagrangian of hadrons interacting with their  
constituent quarks. The coupling strength of the hadrons with the constituent 
quarks is determined by  the compositeness condition~$Z_H=0$  \cite{SWH}
where $Z_H$ is the wave function renormalization constant of the hadron.
The hadron field renormalization constant $Z_H$ characterizes the overlap 
between the bare hadron field and the bound state formed from the 
constituents. Once this constant is set to zero, the dynamics of hadron 
interactions is fully described by constituent quarks in quark loop 
diagrams with local constituent quark propagators.  
Matrix elements are generated by a set of quark loop diagrams 
according to an $1/N_c$ expansion. The ultraviolet divergences of the quark 
loops are regularized by including vertex form factors for the hadron-quark 
vertices which, in addition, describe finite size effects due to the 
non-pointlike structure of hadrons. 
The relativistic constituent quark model contains only a few model parameters:
the light and heavy constituent quark masses and the size 
parameters that describe the size of the distribution of the constituent 
quarks inside the hadron.

In the light quark sector this approach  was successfully applied
to describe the electromagnetic properties
of the pion and the sigma meson  \cite{light_mes},
the electromagnetic form factors and magnetic moments of nucleons, and 
semileptonic decays of the light ground state baryon octet  \cite{light_bar}.
In the heavy quark sector  we have calculated the
baryonic Isgur-Wise functions, decay rates and asymmetry parameters
for  semileptonic decays of baryons,
non-leptonic decay rates, and one-pion and one-photon transitions  
of heavy flavored  baryons \cite{heavy_bar}.
This technique was also applied  
to study the semileptonic decays of the double heavy baryons
and the $B_c$-meson \cite{Bc}. The same bound-state formalism was recently
applied to the analysis of exotic hadron states
--so-called hadronic molecules-- in which the constituents
are hadrons themselves rather than quarks as in the present approach 
\cite{molecules}. 

The local form of the quark propagators used 
in the relativistic constituent quark model can lead to the appearance
of threshold singularities corresponding to free quark production
in transition amplitudes.
As a result, applications of the relativistic quark model had to be restricted
to ground state mesons and baryons with masses less than the sum of the 
constituent quark masses,  and processes with relatively small energies.  
This poses problems for  the description of e.g. light vector mesons, 
e.g. ($\rho$, $K^\ast$), and excited states 
where the particle mass exceeds the sum of the  constituent quarks.

In the present work we propose a refinement of our previous quark model 
approach by effectively implementing quark confinement into the model.
In the  relativistic constituent  quark model  matrix elements
are represented by a set of quark loop diagrams which
are described by a convolution of the local quark propagators and vertex 
functions.
By using Schwinger's $\alpha$-representation for each local quark propagator
and integrating out the loop momenta, one can write the resulting matrix
element expression as an integral which includes integrations 
over a simplex of the $\alpha$-parameters and an integration over
a scale variable extending from zero to infinity. 
By introducing an infrared cutoff on the upper limit of the scale integration 
one can avoid the appearance of singularities in any matrix element. 
The new infrared cutoff parameter $\lambda$ will be taken to have a common 
value for all processes. We determine the parameters of the model by a
fit  to available experimental data.
As a first step,  we apply our approach to evaluate the electromagnetic 
form factors of the pion and the transition form factors of the Dalitz decays
$P \to \gamma \,\,l^+l^-$ and $V \to P \,l^+l^-$. 

The paper is structured as follows.  
In Sec.~\ref{sec:framework}, we discuss the basic notions of 
the relativistic constituent
quark model including a discussion on how gauge invariance is implemented
in the model. We then go on to explain how to effectively implement 
infrared confinement. In Sec.~\ref{sec:fit}  we determine the model parameters 
by a fit  to the decay constants of the $\pi$ and $\rho$ mesons and 
then apply the covariant constituent quark model to evaluate the decays 
$\pi^0 \to \gamma\,\gamma$ and $\rho \to \pi \gamma$ 
including also the form factor behavior of the electromagnetic transitions. 
In Sec.~\ref{sec:extension}, we extend our approach to the strange, charm 
and bottom sectors. We calculate the leptonic decay constants of
both pseudoscalar and vector mesons, and the electromagnetic
and leptonic decay widths.  We calculate
transition form factors and the widths of their Dalitz decays, and compare
the results with recent experimental data. 
Finally, in Sec.~\ref{sec:summary}, we summarize our results. 
In Appendix A we describe in some
technical detail how the one-loop integrations involving tensor loop momenta
have been done. In Appendix B we proof by explicit calculation that
the $\rho \to \gamma$ transition matrix element discussed in the main text
is gauge invariant.

\section{Theoretical framework}
\label{sec:framework}

\subsection{Lagrangian}

The relativistic constituent quark model is based on 
an effective interaction Lagrangian describing the
coupling of hadrons to their constituent quarks.
In this paper we limit ourselves to the meson sector in 
which mesons are described as
quark-antiquark bound states. An extension of the model to baryons 
(three--quark states) and multiquark states is straightforward.  
The coupling of a meson $M(q_1 \bar q_2)$ to its constituent 
quarks $q_1$ and $\bar q_2$ is described by the nonlocal Lagrangian  
\eq
\label{eq:lag}
{\cal L}_{\rm int}^{\rm str}(x) = g_M M (x) \, 
\int\!\! dx_1 \!\! \int\!\! dx_2 
F_M(x,x_1,x_2) \,
\bar q_1(x_1) \, \Gamma_M \, \lambda_M \, q_2(x_2)  \, + \, {\rm h.c.}  
\en 
Here, $\lambda_M$ and   $\Gamma_M$ are Gell-Mann and Dirac matrices
(or a string of Dirac matrices)
chosen appropriately to describe the flavor and spin quantum numbers of the 
meson field $M(x)$.
In the case of pseudoscalar mesons we introduce singlet-octet mixing
with a mixing angle of $\theta_{P} = -18^\circ $, 
while the vector mesons are assumed to be ideally mixed.
The vertex function $F_M(x,x_1,x_2)$ characterizes 
the finite size of the meson.  To satisfy translational invariance the 
vertex function has to obey the identity 
$F_M(x+a,x_1+a,x_2+a) \, = \, F_M(x,x_1,x_2) $
for any given four-vector $a\,$. 
In the following we use a specific form for the vertex function which 
satisfies the above translation invariance relation. One has 
\eq
\label{eq:vertex}
F_M(x,x_1,x_2) \, = \, \delta^{(4)}(x - \sum\limits_{i=1}^2 w_i x_i) \;  
\Phi_M\biggl( (x_1 - x_2)^2 \biggr) 
\en
where $\Phi_M$ is the correlation function of the two constituent quarks 
with masses $m_1$ and $m_2$. 
The variable $w_i$ is defined by $w_i=m_i/(m_1+m_2)$ so that 
$w_1+w_2=1$. 
In principle, the Fourier transform of the correlation function, which we
denote by $\Phi_M(-l^2)$, can be calculated from the solutions
of the Bethe-Salpeter equation for the meson bound  
states~\cite{Roberts:2000aa}.
In Refs.~\cite{light_mes} it was found that, using various
forms for the vertex function, the basic hadron observables
are insensitive to the details of
the functional form of the hadron-quark vertex form factor.
We will use this observation as a guiding principle and choose a simple
Gaussian form for the vertex function $\Phi_M(-l^2)$. 
The minus sign in the argument of $\Phi_{M}(-l^2)$ is chosen to emphasize
that we are working in Minkowski space.
One has 
\eq
\label{eq:fourier}
\Phi_{M}(-l^2) = \exp( l^2/\Lambda_M^2) 
\en 
where the parameter $\Lambda_M$ characterizes the size of the meson. 
Since $l^2$ turns into $-l^2$ in  Euclidean space 
the form~(\ref{eq:fourier}) has the appropriate fall--off
behavior in the Euclidean region. We stress again that any choice for 
$\Phi_M$ is appropriate
as long as it falls off sufficiently fast in the ultraviolet region of
Euclidean space to render the Feynman diagrams ultraviolet finite. 

In the evaluation of the quark-loop diagrams we use
the free local fermion propagator for the constituent quark
\eq
\label{eq:prop}
S_q(k) = \frac{1}{ m_q-\not\! k -i\epsilon }
\en 
with an effective constituent quark mass $m_q$.  The local form
of the quark propagator will lead to the appearance
of threshold singularities corresponding to free quark production.
It is for this reason that we had restricted previous applications of the 
relativistic quark model to the lowest-lying states that satisfy the 
condition
$m_{q_1}+m_{q_2}>m_M$, and to processes with relatively low energies.

\begin{figure}[htbp]
\includegraphics[scale=0.6]{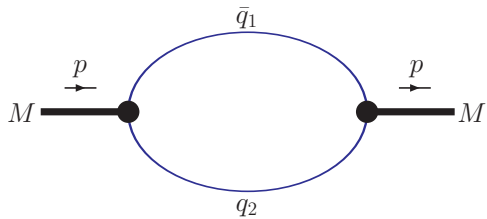}
\caption{Diagram describing the meson mass operator.}
\label{fig:mass}
\end{figure}

The coupling constant $g_M$ in Eq.~(\ref{eq:lag}) is determined by the 
so-called {\it compositeness condition}. 
The compositeness condition requires that the renormalization constant $Z_M$ 
of the elementary meson field $M(x)$ is set to zero, i.e.
\beaa
\label{eq:Z=0}
Z_M=1-\Sigma^\prime_M(m_M^2)=0
\enaa
where $\Sigma^\prime_M(m_M^2)$ is the derivative of the mass operator
corresponding to the self--energy diagram 
Fig.~\ref{fig:mass}. One has  
\eq
\Sigma^\prime_M(m_M^2)=g_M^2\Pi^\prime_M(m_M^2)=
g_M^2\frac{d\Pi_M(p^2)}{dp^2}\big|_{p^2=m_M^2} 
\en 
and where $m_M$ is the meson mass. At this point we take the mesons to be 
spinless for the sake of simplicity. The generalization to mesons 
(or baryons) with arbitrary spin is straightforward.

To clarify the physical meaning of the compositeness condition, we first want 
to remind the reader
that the renormalization constant $Z_M^{1/2}$ can also be interpreted as  
the matrix element between the physical and the corresponding bare state.  
For $Z_M=0$ it then follows that the physical state does not contain  
the bare one and is therefore described as a bound state. 
The interaction Lagrangian Eq.~(\ref{eq:lag}) and 
the corresponding free Lagrangian describe  
both the constituents (quarks) and the physical particles (hadrons)
which are bound states of the constituents.
As a result of the interaction, the physical particle is dressed, 
i.e. its mass and wave function have to be renormalized. 
The condition $Z_M=0$ also effectively excludes
the constituent degrees of freedom from the space of physical states
and thereby guarantees that there will be no double counting.
The constituents exist in virtual states only. 
One of the corollaries of the compositeness condition is the absence 
of a direct interaction of the dressed charged particle with the 
electromagnetic field. Taking into account both the tree-level
diagram and the diagrams with self-energy insertions into the 
external legs (that is the tree-level diagram times $Z_M -1$) yields 
a common factor $Z_M$ which is set to zero. 
This allows for another interpretation of the compositeness condition 
in as much as the  condition $Z_M=0$ leads to the correct normalization of the 
electric form factor of a charged particle at zero momentum transfer. 
By using the Ward identity which relates the electromagnetic vertex 
function at zero momentum transfer to the derivative of the mass operator 
for the on-mass-shell particle and taking into account
the compositeness condition in the form of Eq.~(\ref{eq:Z=0}),
one obtains 
\eq
\label{eq:Z_charge=0} 
\Lambda^\mu(p,p) = \frac{d\Sigma_M(p^2)}{dp_\mu}
                 = 2 p^\mu \frac{d\Sigma_M(p^2)}{dp^2}
                 = 2 p^\mu.
\en
$\Lambda^\mu(p,p)$ is the zero momentum transfer electromagnetic 
vertex function.
The form (\ref{eq:Z_charge=0}) is more suitable for analytical calculations
because the derivative of the mass operator with respect to $p_\mu$
determines the electromagnetic vertex  function at zero momentum transfer.

\subsection{Inclusion of photons}

The interaction with the electromagnetic field is introduced
in two stages. The free Lagrangian of quarks and hadrons
is gauged in the standard manner by using minimal
substitution:

\begin{equation}
\label{eq:photon}
\partial^\mu M^\pm \to (\partial^\mu \mp ie A^\mu) M^\pm ,
\hspace{1cm} 
\partial^\mu q \to (\partial^\mu - ie_q A^\mu) q, 
\hspace{1cm} 
\partial^\mu \bar q \to (\partial^\mu +ie_q A^\mu)\bar q,
\end{equation}
where $e$ is the positron (or proton) charge and where $e_q$ is the 
quark's charge ($e_u= \tfrac23\, e$, $e_d=-\,\tfrac13\, e$, etc.). 
Minimal substitution gives us the 
first piece of the electromagnetic interaction Lagrangian  
\eq\label{eq:L_em_1}
{\cal L}^{\rm em (1)}_{\rm int}(x) &=& 
\sum_q e_q\, A_\mu(x)\,J^\mu_q(x) + e\, A_\mu(x)\,J^\mu_M(x)
+ \, e^2  A^2(x) M^-(x) M^+(x) \,,
\nonumber\\
 J^\mu_q(x) &=& \bar q(x) \gamma^\mu q(x), \qquad 
J^\mu_M(x) = 
i\,\Big(M^-(x)\partial^{\,\mu} M^+(x) - M^+(x)\partial^{\,\mu} M^-(x)\Big).
\en 
It is important to reiterate that  there is no direct coupling of the photon
to the meson in our relativistic quark model due to the compositeness condition
$Z_M=0$ as has been emphasized in Sec.~\ref{sec:framework}~A. 

Gauging the nonlocal piece of the Lagrangian in Eq.~(\ref{eq:lag})
proceeds in a way suggested in~\cite{gauging}.
In order to guarantee local gauge invariance of the strong interaction 
Lagrangian, one multiplies each quark field $q(x_i)$ in  
${\cal L}_{\rm int}^{\rm str}$ with a gauge field exponential
according to
\eq 
\label{eq:gauging}
{\cal L}_{\rm int}^{\rm str + em(2)}(x) =
g_M M(x)\int\!\! dx_1 \!\!\int\!\!dx_2 F_M (x,x_1,x_2)
\bar q_1(x_1)\, e^{ie_{q_1} I(x_1,x,P)} \,\Gamma_M \, \lambda_M\,  
e^{-ie_{q_2} I(x_2,x,P)}\,  q_2(x_1),   
\en 
where
\eq 
\label{eq:path}
I(x_i,x,P) = \int\limits_x^{x_i} dz_\mu A^\mu(z). 
\en 
It is readily seen that the full Lagrangian is invariant 
under the local gauge transformations
\eq 
q_i(x) &\to& e^{ie_{q_i} f(x)} q_i(x),    \hspace{1cm}
\bar q_i(x) \to \bar q_i(x) e^{-ie_{q_i} f(x)}, \hspace{1cm}
M(x)\to e^{ie_{M} f(x)}M(x)\,, \nonumber\\
A^\mu(x)&\to& A^\mu(x)+\partial^\mu f(x) \,,
\en
where $e_M=e_{q_2}-e_{q_1}$ is the meson electric charge. 

The second term of the electromagnetic interaction Lagrangian
${\cal L}^{em}_{\rm int; 2}$ arises when one expands the gauge exponential
in powers of $A_\mu$ up to the order of perturbation 
theory that one is considering. 
Superficially the results appear to depend on the path $P$
which connects the end-points in the path integral in Eq~(\ref{eq:path}).
However, one needs to know only derivatives of the path
integrals when doing the perturbative expansion.
One can make use of the formalism developed in~\cite{gauging}
which is based on the path-independent definition of the derivative of 
$I(x,y,P)$: 
\begin{eqnarray}\label{eq:path1}
\lim\limits_{dx^\mu \to 0} dx^\mu 
\frac{\partial}{\partial x^\mu} I(x,y,P) \, = \, 
\lim\limits_{dx^\mu \to 0} [ I(x + dx,y,P^\prime) - I(x,y,P) ]
\end{eqnarray}
where the path $P^\prime$ is obtained from $P$ by shifting the end-point $x$
by $dx$.
Use of the definition (\ref{eq:path1}) 
leads to the key rule
\begin{eqnarray}\label{eq:path2}
\frac{\partial}{\partial x^\mu} I(x,y,P) = A_\mu(x)
\end{eqnarray}
which states that the derivative of the path integral $I(x,y,P)$ does 
not depend on the path $P$ originally used in the definition. The non-minimal 
substitution (\ref{eq:gauging}) is therefore completely equivalent to the 
minimal prescription as is evident from the identities (\ref{eq:path1}) or 
(\ref{eq:path2}). 
The method of deriving Feynman rules for the non-local coupling 
of hadrons to photons and quarks was already developed in 
Refs.~\cite{light_mes} 
and will be discussed in the next section 
where we apply the formalism to the physical processes considered in this
paper. 

\begin{figure}[htbp]
\includegraphics[scale=0.5]{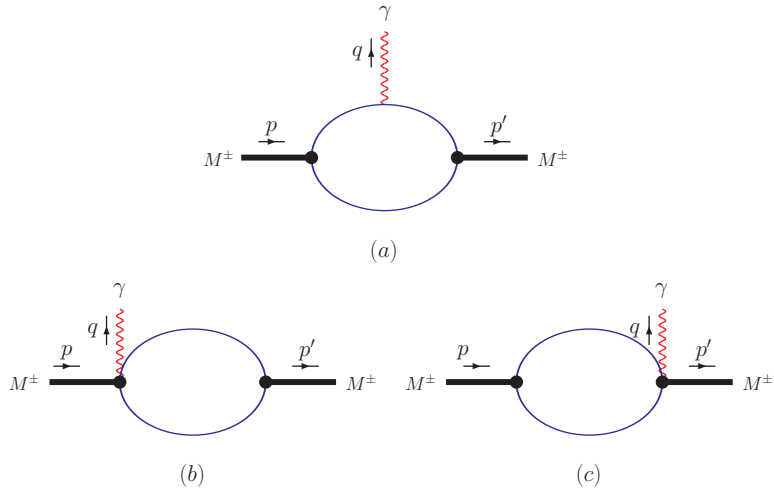}
\caption{Diagrams describing 
 meson electromagnetic vertex function.} 
\label{fig:MMg}
\end{figure}

For example, expanding the Lagrangian Eq.~(\ref{eq:gauging}) up to the first
order in $A^{\mu}$ one obtains $(l = w_1 p_1 + w_2 p_2 )$ 
\begin{eqnarray}\label{eq:first}
{\cal L}_{\rm int}^{\rm em(2)}(x) &=&   
g_M M(x)\int\!\! dx_1 \!\!\int\!\!dx_2\int\!\!dy\, 
E^\mu_M(x,x_1,x_2,y)\,  A_\mu(y)\, 
\bar q_1(x_1)\Gamma_M\lambda_M q_2(x_2)\,,
\nonumber\\ 
&&\nonumber\\ 
E^\mu_M(x,x_1,x_2,y) &=&
\int\frac{dp_1}{(2\pi)^4}\int\frac{dp_2}{(2\pi)^4}
\int\frac{dq}{(2\pi)^4}
e^{ip_1(x_1-x)-ip_2(x_2-x)+iq(y-x)}E^\mu_1(p_1,p_2,q)\,, \\
&&\nonumber\\ 
E^\mu_1(p_1,p_2,q) &=& -e_{q_1}w_{1}(w_1 q^\mu+2 l^\mu)
\int\limits_0^1 dt \Phi'_H\left(-t(w_1 q+l)^2-(1-t)l^2\right)
\nonumber\\
&+&
e_{q_2}w_2(w_2 q^\mu-2  l^\mu)
\int\limits_0^1 dt \Phi'_M\left(-t(w_2 q-l)^2-(1-t)l^2\right)\,,
\nonumber
\end{eqnarray}
Let us emphasize that the vector Ward-Takahashi identity for an off-shell 
photon is also satisfied in this approach: 
\bea
q_\mu\Lambda^\mu(p,p^\prime)=\Sigma_M(p^2)-\Sigma_M(p^{\prime\,2})\,, 
\ena 
where $\Lambda^\mu(p,p^\prime)$ is the meson electromagnetic vertex function 
described by the diagrams shown in Fig.~\ref{fig:MMg}. 
In order to guarantee the universality of neutral and charged current
transitions (conserved vector current hypothesis) one also has to gauge
the Lagrangian with regard to the electroweak interactions. This has been
described in detail in Ref.~\cite{light_bar} (see, also Sec.~\ref{sec:fit}).  

\subsection{Infrared confinement}
\label{Sec:Cf} 

Let us reiterate that the relativistic constituent quark model as described
up to this point has a limited range of applications due to the threshold 
constraint $m_1 + m_2 > m_M$.
In particular, processes involving light vector mesons,
e.g. the $\rho$ and $K^\ast$,  and excited states  
cannot be treated within the model. 
In this section we extend the applicability of the relativistic 
constituent quark model by taking into account quark confinement effects. 
In order to elucidate how confinement is implemented
in our framework we begin by considering a scalar one-pole propagator. 
The Schwinger representation of the propagator reads 
\bea
\label{eq:prop_1}
\frac{1}{m^2-p^2}=
\int\limits_0^\infty d\alpha\exp[-\alpha(m^2-p^2)] 
\ena 
where the integration over the Schwinger parameter runs from 0 to $\infty$.
Instead of integrating to infinity we introduce an upper integration limit 
$1/\lambda^2$. We call the dimensional parameter $\lambda$ (with mass
dimension $[m]$) the infrared 
confinement scale. By means of the cutoff one obtains an entire function 
which can be interpreted as a confined propagator, 
\bea
\label{eq:prop_2}
\int\limits_0^{1/\lambda^2} d\alpha\exp[-\alpha(m^2-p^2)] 
= \frac{1 - \exp[-(m^2-p^2)/\lambda^2]}{m^2-p^2} \,.
\ena 
Similar ideas have also been pursued in Refs.~\cite{NJL} where an infrared 
cutoff had been introduced in the context of a 
Nambu-Jona-Lasinio model. 
Note that the propagator for a particle in a constant self-dual field
has a form similar to  Eq.~(\ref{eq:prop_2}). Such vacuum gluon configurations
have been studied in \cite{Leutwyler:1980ev} and were then employed
e.g. in~\cite{Efimov:1995uz} to construct a model with
confined constituent quarks.
The propagator in  Eq.~(\ref{eq:prop_2}) does not have any singularities 
in the finite $p^2$-plane,  thus indicating the absence of a single quark 
in the asymptotic space of states. 

However, the use of confined propagators in the form of entire functions
has its own difficulties. 
The convolution of entire functions leads to a rapid growth of 
physical matrix elements once the hadron masses
and energies of the reaction have been fixed. The numerical
results become very sensitive to changes of the input
parameters which requires extreme fine-tuning. For these reasons, we suggest 
to proceed in the following way.

Let us consider a general $l$-loop Feynman diagram with $n$ 
propagators. One writes, using again the Schwinger parameterization, 
\eq
\Pi(p_1,\ldots,p_n) = \int\limits_0^\infty d^n \alpha 
\int [d^4 k]^l \, \Phi \, \exp[-\sum\limits_{i=1}^n \alpha_i (m_i^2-p_i^2)] \, 
\label{eq:loop_1}
\en 
where $\Phi$ stands for the numerator product of propagators and vertex 
functions. After doing the loop integrations one obtains
\eq
\Pi =  \int\limits_0^\infty d^n \alpha \, F(\alpha_1,\ldots,\alpha_n) \,,
\en
where $F$ stands for the whole structure of a given diagram. 
The set of Schwinger parameters $\alpha_i$ can be turned into a simplex by 
introducing an additional $t$--integration via the identity 
\eq 
1 = \int\limits_0^\infty dt \, \delta(t - \sum\limits_{i=1}^n \alpha_i)
\en 
leading to 
\eq
\label{eq:loop_2} 
\Pi   = \int\limits_0^\infty dt t^{n-1} \int\limits_0^1 d^n \alpha \, 
\delta\Big(1 - \sum\limits_{i=1}^n \alpha_i \Big) \, 
F(t\alpha_1,\ldots,t\alpha_n)\,. 
\en
As in Eq.~(\ref{eq:prop_2}) we cut off the upper integration at $1/\lambda^2$
and obtain
\eq
  \Pi^c =   
\int\limits_0^{1/\lambda^2} dt t^{n-1} \int\limits_0^1 d^n \alpha \, 
\delta\Big(1 - \sum\limits_{i=1}^n \alpha_i \Big) \, 
F(t\alpha_1,\ldots,t\alpha_n)
\en  
By introducing the infrared cutoff one has removed all possible thresholds 
in the quark loop diagram. We take the cutoff parameter $\lambda$ to be the 
same in all physical processes. 

In order to make contact with recent ideas on the holographic description
of particle interactions we change the integration variable $t$ in 
Eq.~(\ref{eq:loop_2}) to $z$ with $t=z^2$. One can then interpret
$z$ as an extra space coordinate
and the upper integration limit  $z_{\rm IR}=1/\lambda$ as
the infrared scale where quarks are confined and hadronized. 
The analogy to the extra dimension holographic coordinate $z$ introduced 
in holographic model is now apparent. We mention that the holographic model 
of hadrons is motivated by 
the anti-de Sitter/conformal field theory (AdS/CFT) correspondence, 
see Refs.~\cite{AdS}. There the truncation over the holographic coordinate 
$z$ is necessary in order to break conformal invariance and to incorporate 
confinement in the infrared region. 
 
As a further illustration of the infrared confinement effect relevant to the 
applications in this paper 
we consider the case of a scalar one--loop two--point 
function. One has
\eq 
\Pi_2(p^2) = \int \frac{d^4k_E}{\pi^2} 
\frac{e^{-sk_E^2}}{[m^2 + (k_E+\frac{1}{2}p_E)^2]
[m^2 + (k_E-\frac{1}{2}p_E)^2]} 
\en 
where the numerator factor $e^{-sk_E^2}$ comes from the product of nonlocal 
vertex form factors 
of Gaussian form; $k_E$, $p_E$ are Euclidean momenta. Doing the loop 
integration one obtains 
\eq 
\Pi_2(p^2) = \int\limits_0^\infty dt \frac{t}{(s+t)^2} 
\int\limits_0^1 d\alpha \, \exp\Big\{ - t[m^2 - \alpha(1-\alpha)p^2] 
+ \frac{st}{s+t} \Big(\alpha - \frac{1}{2}\Big)^2 p^2 \Big\} \,.  
\en 
The integral $\Pi_2(p^2)$ can be seen to have a branch point at $p^2=4m^2$. 
By introducing a cutoff in the $t$--integration one obtains  
\eq 
\Pi_2^c(p^2) = \int\limits_0^{1/\lambda^2} dt \frac{t}{(s+t)^2} 
\int\limits_0^1 d\alpha \, \exp\Big\{ - t[m^2 - \alpha(1-\alpha)p^2] 
+ \frac{st}{s+t} \Big(\alpha - \frac{1}{2}\Big)^2 p^2 \Big\} \,,   
\en 
where the one--loop two--point function $\Pi_2^c(p^2)$ no longer has 
a branch point at $p^2=4m^2$. 

Such a confinement scenario can be realized with only minor changes in 
our approach by shifting the upper $t$--integration limit from 
infinity to $1/\lambda^2$. The confinement scenario also allows  
to include all possible resonance states in our calculations. 
First calculations done in this paper show that the 
limited set of adjustable parameters of the model (size parameters, 
constituent quark masses and the confinement scale $\lambda$) leads to 
a consistent description of a large number of low energy mesonic processes. 
We envisage a multitude of further applications as e.g. in the baryon sector.

\section{Basic properties of \boldmath{\Large $\pi$} and 
\boldmath{\Large $\rho$} mesons}
\label{sec:fit}

\begin{figure}[htbp]
\begin{center}
\hspace*{-0.5cm}
\begin{tabular}{c}
\includegraphics[scale=0.5]{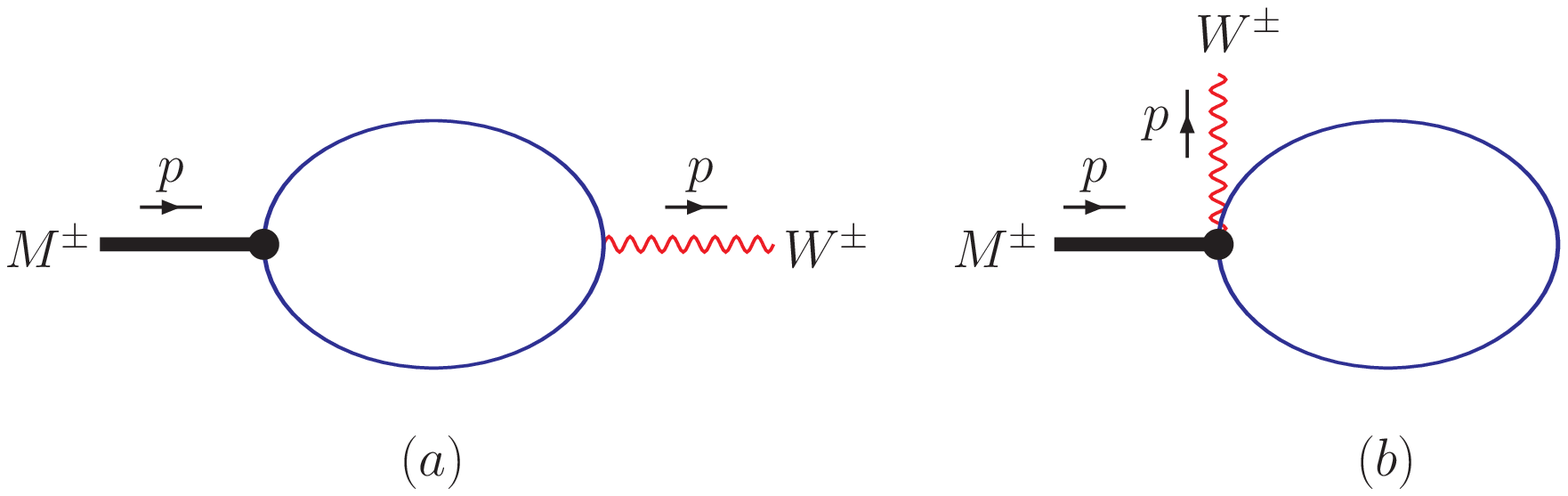} \\
\includegraphics[scale=0.5]{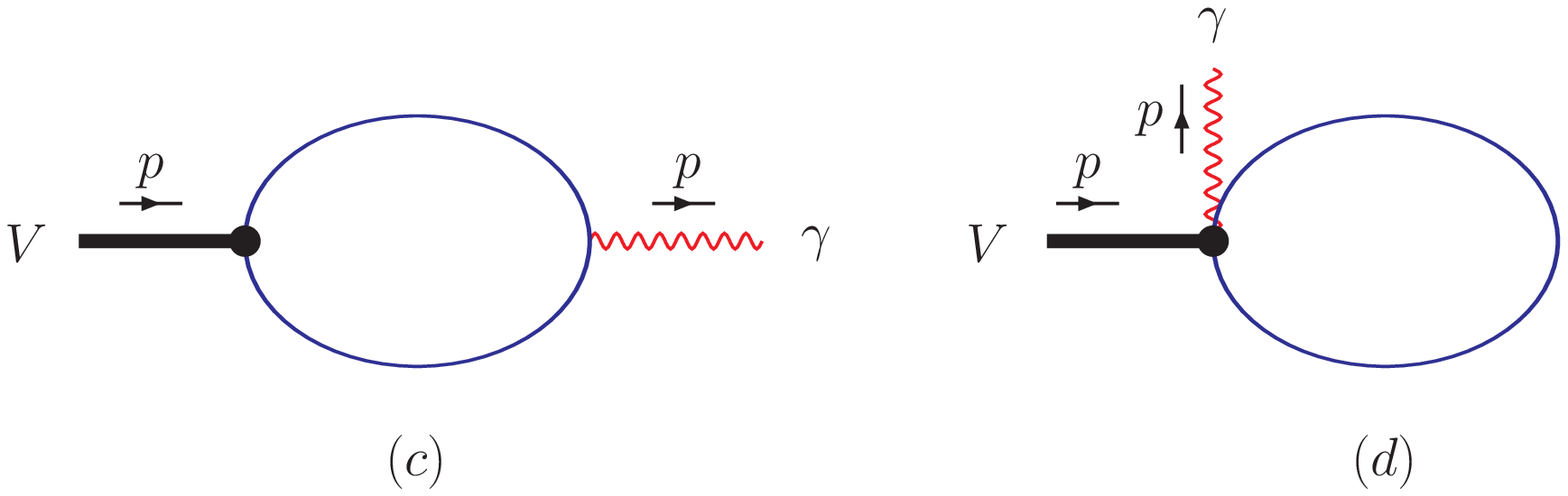}
\end{tabular}
\end{center}
\caption{
Diagrams describing $M^\pm \to W^\pm $ (upper panel) and
$V \to \gamma$ (lower panel) transitions.
}
\label{fig:MW}
\end{figure}

\begin{figure}[htbp]
\includegraphics[scale=0.5]{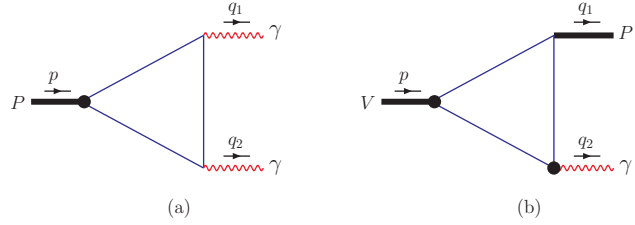}
\caption{Diagrams describing $P\to\gamma\gamma$ and $V\to P\gamma$ 
transitions.}
\label{fig:VPg}
\end{figure}

In this section we discuss applications of the relativistic 
constituent quark model including infrared confinement to the decays 
of the $\pi$ and $\rho$ mesons.
We start by fitting the model parameters.
We calculate  the leptonic constants $f_\pi$, $g_{\rho\gamma}$ and 
the electromagnetic couplings $g_{\pi\gamma\gamma}$ and $g_{\rho\pi\gamma}$. 
The relevant quark model diagrams are shown in 
Figs.~\ref{fig:MW} and \ref{fig:VPg}.
The corresponding Feynman one-loop integrals read: 
\bea
%
%
f_\pi\,  p^\mu &=& N_c g_\pi\! \int\!\!\frac{d^4k}{(2\pi)^4i} 
\Big\{  
\Phi_\pi(-k^2)\,
\text{tr}\,\Big[O^\mu S(k+\tfrac12\, p)\gamma^5 S(k-\tfrac12\, p)\Big]
\nonumber\\
&&
+\, \int\limits_0^1 d\alpha \Phi'_\pi(-z_\alpha)
(2\,k+\tfrac12\, p)^\mu \text{tr}\Big[S(k)\Big]
\Big\},
\label{fpi}\\ 
&& z_\alpha= \alpha (k+\tfrac12\, p)^2 + (1-\alpha)k^2 ,
\nonumber\\[2ex] 
%
%
g_{\rho\gamma}\,  \big(g^{\mu\nu}p^2-p^\mu p^\nu\big) &=&
\frac{N_c  g_\rho }{\sqrt{2}}\!\int\!\!\frac{d^4k}{(2\pi)^4i} 
\Big\{   
\Phi_\rho(-k^2)\,
\text{tr}\Big[\gamma^\mu S(k+\tfrac12\, p)\gamma^\nu S(k-\tfrac12\, p)\Big]
\nonumber\\
&&
-\,\int\limits_0^1 d\alpha \Phi'_\rho(-z_\alpha) 
(2\,k+\tfrac12\, p)^\mu \text{tr}\Big[\gamma^\nu S(k)\Big] 
\Big\},
\label{VGint} \\[2ex] 
%
%
g_{\pi\gamma\gamma}\,  \epsilon^{\mu\nu q_1 q_2}&=&
\frac{\sqrt{2} N_c}{3}\!\int\!\!\frac{d^4k}{(2\pi)^4i} 
\Phi_\pi(-k^2)\,
\text{tr}\,\Big[i\gamma^5S(k+\tfrac12\, p)\gamma^\mu 
S(k-\tfrac12\, p)\gamma^\nu S(k+\tfrac12\, p - q_1)\Big],
\label{Pgg} \\[2ex]
%
%
g_{\rho\pi\gamma}\, \epsilon^{\mu\nu q_1 q_2 } &=& 
\frac{N_c g_\rho g_\pi}{3}\!\int\!\!\frac{d^4k}{(2\pi)^4i} 
\Phi_\rho(-k^2)\Phi_\pi\Big(-(k+\tfrac12\, q_2)^2\Big)
\nonumber\\
&&
\times\,
\text{tr}\,\Big[i\gamma^5 S(k+\tfrac12\, p)\gamma^\mu S(k-\tfrac12\, p)
\gamma^\nu S(k+\tfrac12\, p - q_1)\Big],
\label{VPg} 
\ena 
where $N_c=3$ is the number of colors,
$O^\mu=\gamma^\mu(I-\gamma^5)$ is the left-chiral weak coupling matrix
and the local propagators $S(k)$ etc. are defined in Eq.~(\ref{eq:prop}).

The meson-quark coupling constants are determined from
the compositeness condition Eq.~(\ref{eq:Z_charge=0}). This requires the
evaluation of the derivative of the mass operator.
For the pseudoscalar and vector mesons treated in this paper the
derivatives of the mass operators read
\begin{eqnarray}
\Pi'_P(p^2)&=& 
\frac{1}{2p^2} p^\alpha\frac{d}{dp^\alpha} 
N_c\int\!\!\frac{d^4k}{(2\pi)^4i} \Phi^2_P(-k^2)
{\rm tr} \biggl[\gamma^5 S_1(k+w_1 p) \gamma^5
                         S_2(k-w_2 p) \biggr]
\nonumber\\[2ex]
&=&\frac{N_c}{2p^2}\int\!\!\frac{d^4k}{(2\pi)^4i} \Phi^2_P(-k^2)
{\rm tr} \biggl[\gamma^5 S_1(k+w_1 p)\,
        w_1\!\!\not\!p\, S_1(k+w_1 p) \gamma^5
                         S_2(k-w_2 p) \biggr]
 + (m_1\leftrightarrow m_2),
\label{eq:PS}\\[4ex]
\Pi'_V(p^2)&=& 
\frac{1}{3}\biggl[g^{\mu\nu}-\frac{p^\mu p^\nu}{p^2}\biggr] 
\frac{1}{2p^2} p^\alpha\frac{d}{dp^\alpha}
N_c\int\!\! \frac{d^4k}{(2\pi)^4i} \Phi^2_V(-k^2) 
{\rm tr} \biggl[\gamma^\mu S_1(k+w_1 p) \gamma^\nu
                           S_2(k-w_2 p)\biggr]
\nonumber\\[2ex]
&=&
\frac{1}{3}\biggl[g^{\mu\nu}-\frac{p^\mu p^\nu}{p^2}\biggr] 
\frac{N_c}{2p^2}\int\!\! \frac{d^4k}{(2\pi)^4i} \Phi^2_V(-k^2) 
{\rm tr} \biggl[\gamma^\mu S_1(k+w_1 p)\, 
         w_1\!\! \not\!p\, S_1(k+w_1 p) \gamma^\nu
                           S_2(k-w_2 p)\biggr]
\nonumber\\[2ex]
&& +\, (m_1\leftrightarrow m_2).
\label{eq:V}
\end{eqnarray}
Because of the the $p^\mu$-derivative Eqs.~(\ref{eq:PS}) and (\ref{eq:V})
contain three propagator factors.
The evaluation of the one-loop integrals Eqs.~(\ref{fpi})--(\ref{eq:V})
proceeds as described in Appendix A. 

There are four adjustable parameters:
the constituent quark mass  $m\equiv m_{u(d)}$,
the size parameters $\Lambda_\pi$ and $\Lambda_\rho$,
and the scale  parameter $\lambda$ characterizing the  infrared confinement.
A least square fit to the observables  yields the fit parameters 
\eq 
\begin{array}{ccccc} 
 \,\,\,  m  \,\,\,&  \,\,\,\Lambda_\pi \,\,\, & 
 \,\,\, \Lambda_\rho  \,\,\, & \,\,\,  \lambda \,\,\, &  \\
\hline 
\,\,\,0.217\,\,\, & \,\,\,0.711\,\,\, &\,\,\, 0.295\,\,\, 
& \,\,\,0.181 \,\,\,&  {\rm GeV} \\
\end{array}\;.
\en 
In Table~\ref{table:basic_fit} we compare the results of the fit with 
available experimental data. 
\begin{table}[htpb]
\renewcommand{\arraystretch}{1.6}
     \setlength{\tabcolsep}{0.5cm}
     \centering
     \caption{Basic properties of the $\pi$ and $\rho$ meson. 
\label{table:basic_fit}}
\begin{tabular}{lcc}
\hline
\hline 
Quantity & Our & Data~\cite{Amsler:2008zz}\\ 
\hline\hline
$f_\pi$, MeV        & 130.2 & $130.4\pm 0.04 \pm 0.2$\\ 
\hline
$g_{\pi\gamma\gamma}$, 
GeV$^{-1}$          & 0.23 & 0.276 \\ 
\hline 
$g_{\rho\gamma}$    & 0.2 & 0.2 \\ 
\hline 
$g_{\rho\pi\gamma}$, 
GeV$^{-1}$          & 0.75 & 0.723 $\pm$ 0.037\\ 
\hline\hline
\end{tabular}
\end{table} 

As a first application of our approach we calculate
the pion electromagnetic form factor $F_\pi$ generated by the diagrams 
in Fig.~\ref{fig:MMg}, and the pion transition form factor
$F_{\pi\gamma\gamma^\ast}$  generated by the diagram in Fig.~\ref{fig:VPg}(a). 
In the first case, we are interested in the space-like region $q^2=-Q^2$.
In the second case one photon 
is on-mass-shell $q_1^2=0$ and the second photon has 
a space-like momentum squared $q_2^2=-Q^2$.
The electromagnetic radii are related to the slope of 
form factors at the origin  $ r^2 = - 6 F'(0)$.
Our result for the electromagnetic radius  $r_{\pi} = 0.612 \ {\rm fm}$
is in good agreement with  the present world average data of 
$r_{\pi} = (0.672 \pm 0.008)$ fm~\cite{Amsler:2008zz}.   
The result for the radius of the transition
form factor $r^2_{\pi\gamma}  = 0.315 \ {\rm fm}^2$ 
confirms the monopole-type approximation of 
the CLEO data~\cite{Gronberg:1997fj} 
and is close to the CELLO measurement~\cite{Behrend:1990sr} of
$r^2_{\pi\gamma}  = 0.42 \pm 0.04$ fm$^2$. 

\begin{figure}[htbp]
\begin{center}
\hspace*{-0.5cm}
\begin{tabular}{lr}
\includegraphics[scale=0.7]{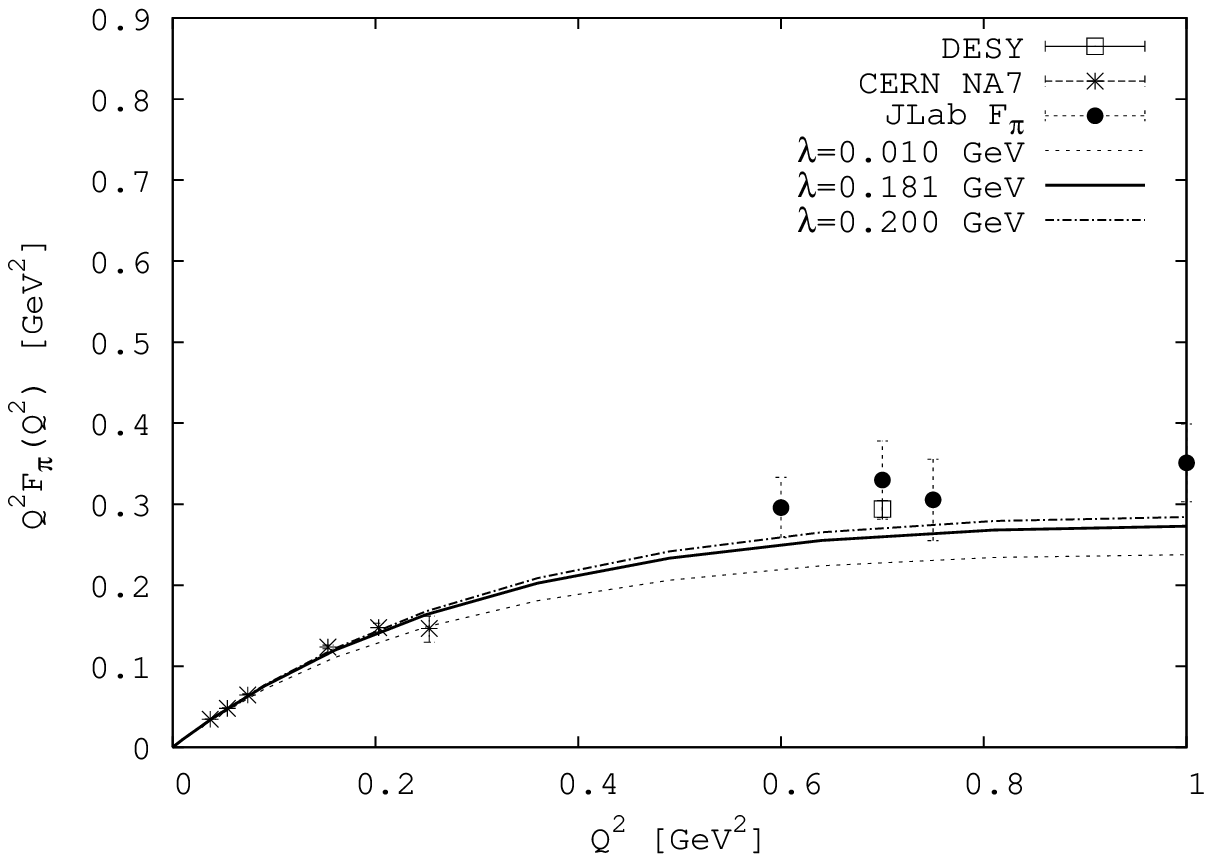} &
\includegraphics[scale=0.7]{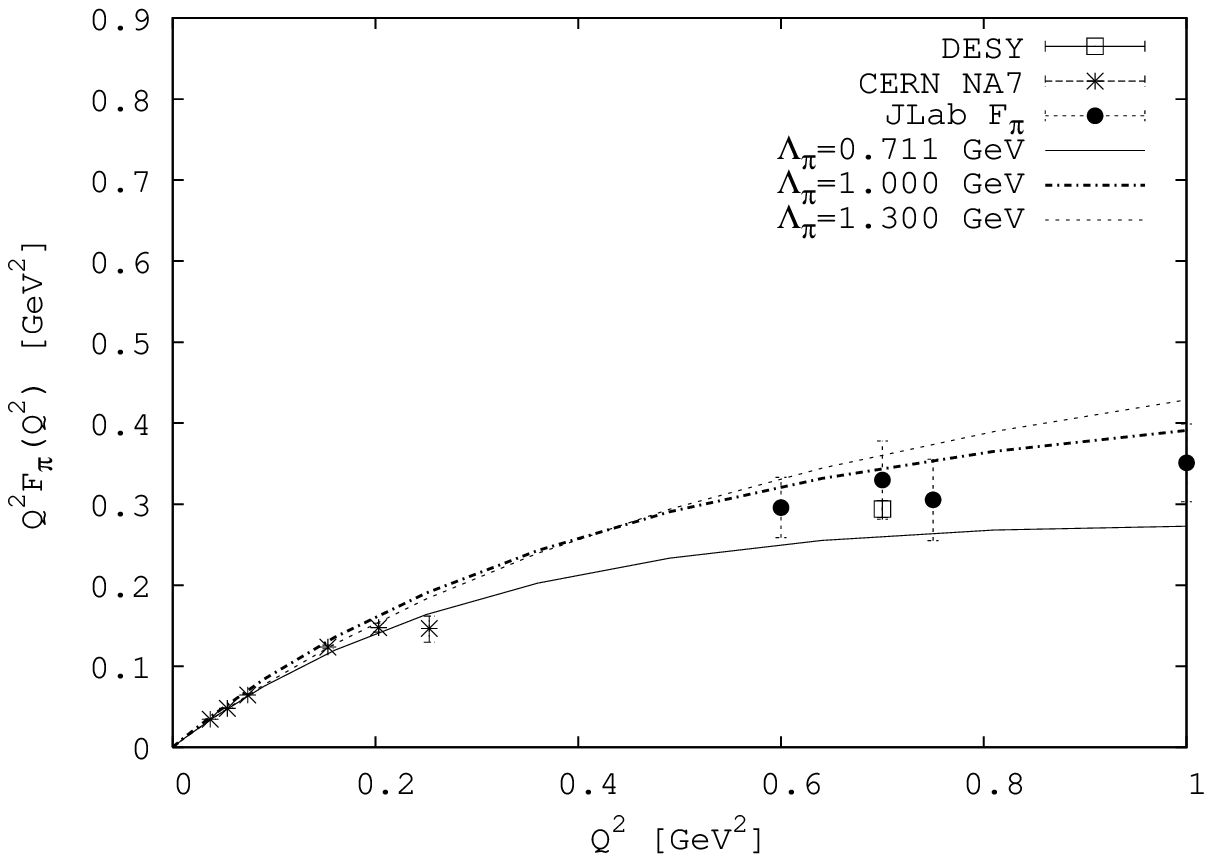}
\end{tabular}
\end{center}
\caption{Form factor $\pi^\pm\to\pi^\pm\gamma^\ast$ as function 
of the space-like photon momentum $Q^2$. Data are taken from 
the JLAB $F_\pi$ Collaboration~\cite{Volmer:2000ek}, DESY~\cite{Brauel:1979zk} 
and CERN NA7 Collaboration~\cite{Amendolia:1986wj}.}
\label{fig:Fpi}
\end{figure}

\begin{figure}[htbp]
\begin{center}
\hspace*{-0.5cm}
\begin{tabular}{lr}
\includegraphics[scale=0.7]{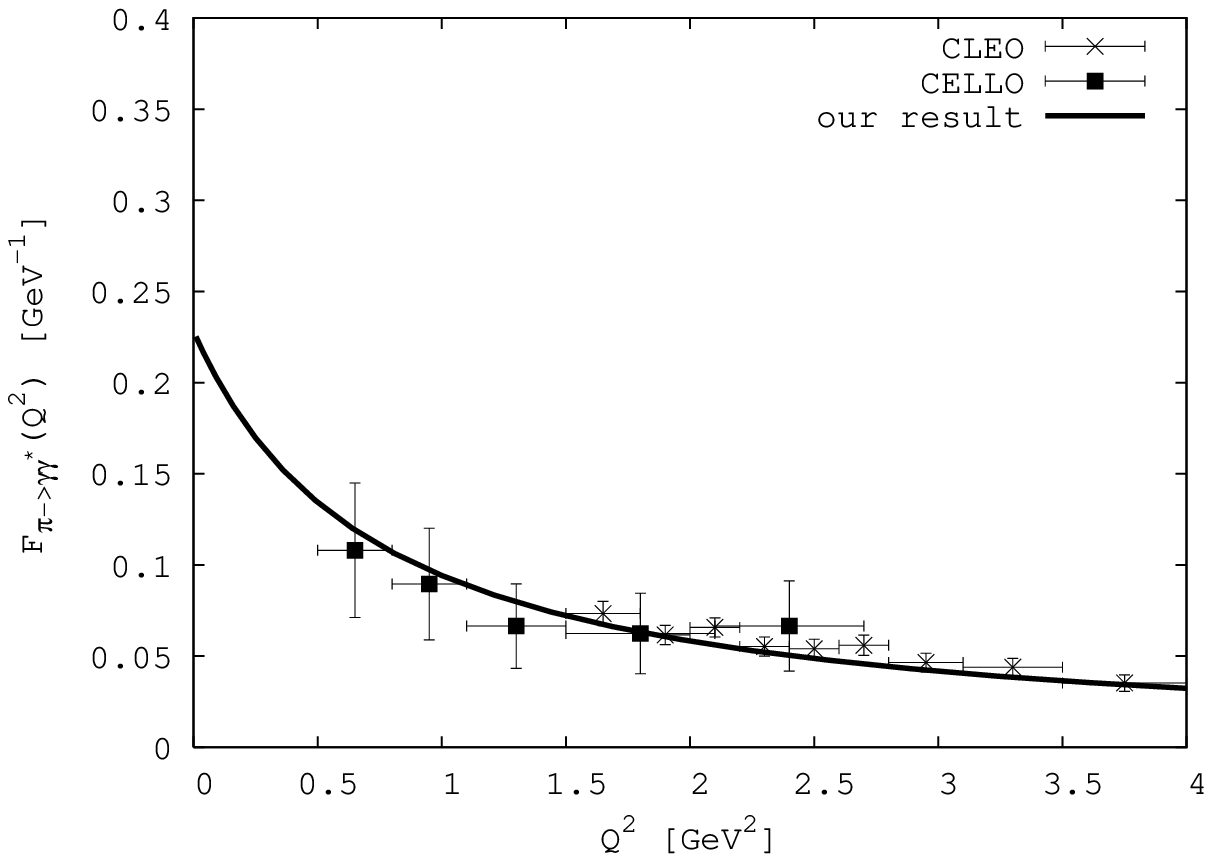} &
\includegraphics[scale=0.7]{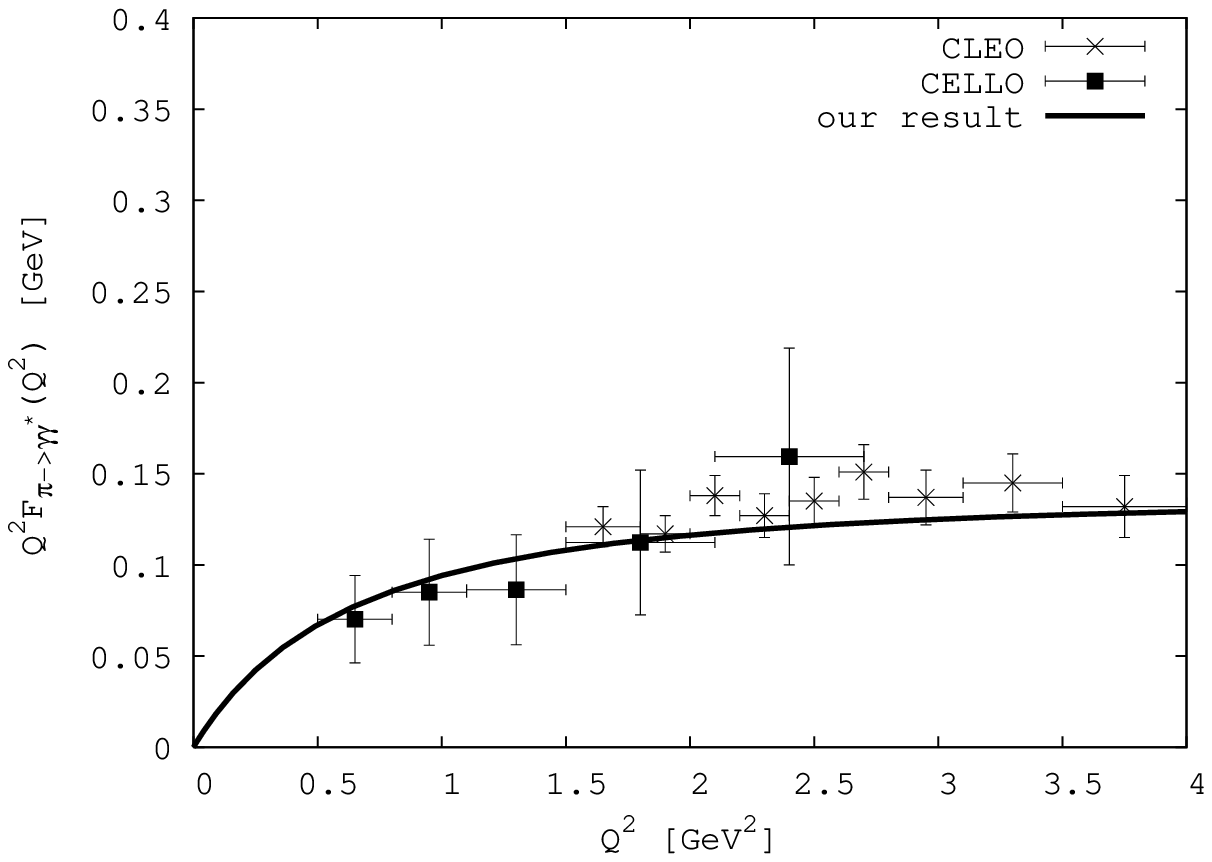}
\end{tabular}
\end{center}
\caption{Form factor $\gamma\gamma^\ast \to \pi^0$ as function of the 
space-like photon momentum $Q^2$. Data are taken from 
CLEO \cite{Gronberg:1997fj}.} 
\label{fig:Pgg}
\end{figure}

The electromagnetic pion transition form factor is displayed in 
Fig.~\ref{fig:Fpi} and compared with data from DESY~\cite{Brauel:1979zk}, 
the Jefferson Lab $F_\pi$ Collaboration~\cite{Volmer:2000ek} 
and the CERN NA7 Collaboration~\cite{Amendolia:1986wj}. 
In Fig.~\ref{fig:Pgg}  we display the  $F_{\pi\gamma\gamma^\ast}(Q^2)$ 
form factor in the space-like region $Q^2$ up to 4 GeV$^2$. 

Note that the slopes of the theoretical curves are quite sensitive 
to variations  of the size parameter $\Lambda_\pi$.
In order to exhibit the sensitivity to $\Lambda_\pi$ we plot three curves
for $\Lambda_\pi = 0.711, 1$ and $1.3$ GeV in Fig.~\ref{fig:Fpi}. Similarly, 
we display the sensitivity of the results on the infrared confinement 
scale for values of $\lambda = 0.010, 0.181$ and $0.200$ GeV. 

One should mention that our $F_{\pi\gamma\gamma^\ast}(Q^2)$ form factor 
behaves as  $1/Q^2$ at large $Q^2$ in accordance with
perturbative QCD. The calculated form factor behavior disagrees 
with the new data above  4 GeV$^2$ presented by 
the {\it BABAR} Collaboration~\cite{Aubert:2009mc}. 
For a recent theoretical analysis of the {\it BABAR} data, see 
e.g. Refs.~\cite{Dorokhov:2009dg}.


\section{Extension to other mesons and some applications}
\label{sec:extension}

\subsection{An extension to strange, charm and bottom flavors} 

In this subsection we extend our approach to mesons containing 
strange, charm and bottom quarks. We accordingly have to introduce a new set 
of fit parameters, namely, the values of constituent quark masses
$m_q$ $(q=s,c,b)$ and the values of size parameters $\Lambda_M$ for the 
corresponding mesons.  Note that we keep the value of the confinement 
scale parameter $\lambda$ fixed for all physical processes.
For the fitting procedure we choose the weak and electromagnetic
leptonic  decay constants, see, Tables~\ref{tab:weak}-\ref{tab:widths}. 
We use the electromagnetic leptonic decay constants
for the neutral vector mesons defined by  $f_V = m_V g_{V\gamma}$
where the dimensionless constant $g_{V\gamma}$ was introduced
in Sec.~\ref{sec:fit} by Eq.~(\ref{VGint}). 
Some comments should be done with respect to our fitting procedure.
The values of strange, charm and bottom
quark masses and of the size parameters of hadrons
are determined by a least square fit to available experimental data
for weak and electromagnetic leptonic decay constants
shown in Tables~\ref{tab:weak} and \ref{tab:em}.
In addition, we include the decays $\eta_c\to\gamma\gamma$
and $J/\psi\to\eta_c\gamma$ (see Table~\ref{tab:widths})
for a better determination of the charm parameters. Since
the size parameters $\Lambda_{J/\psi}$  and  $\Lambda_{\eta_c}$
are found  to be close to each other we assume that the unknown
value of $\Lambda_{\eta_b}$ is approximately the same as
$\Lambda_{\Upsilon}$.
The description of the $\eta$ and $\eta'$ mesons is more involved
because both contain a non-strange and strange quark-antiquark
admixture which is characterized by the mixing angle $\theta_P$.
Due to SU(3)-breaking we distinguish between the size parameters
of the non-strange and strange pieces of the quark current, i.e.
\eq
\eta &\longrightarrow&
-\,\tfrac{1}{\sqrt{2}}\sin\delta\, \Phi_{\Lambda_{\eta}}(\bar u u +\bar d d)
-\,\cos\delta\, \Phi_{\Lambda_{\eta_s}}\,\bar s s \,,
\nonumber\\
\eta' &\longrightarrow&
+\,\tfrac{1}{\sqrt{2}}\cos\delta\, \Phi_{\Lambda_{\eta'}}\,(\bar u u +\bar
d d)
-\,\sin\delta\, \Phi_{\Lambda_{\eta'_s}}\, \bar s s \,,
\nonumber\\
\delta &=& \theta_P-\theta_I, \qquad \theta_I=\arctan{\tfrac{1}{\sqrt{2}}}.
\label{eq:mixing}
\en
We use the experimental data of the electromagnetic decays
involving $\eta$ and $\eta'$ (see Table~\ref{tab:widths})
to fit the size parameters and the mixing angle. It appears that
the best value for the mixing angle $\theta_P$ is $-18^o$. This result
is consistent with the range of values between $-10^o$ and  
$-20^o$~\cite{Amsler:2008zz} obtained
in direct extractions of $\theta_P$ from decay data involving $\eta$  
and $\eta'$.
A mixing angle of or very close to  $-18^o$ is also found in a direct analysis
of the ratio 
$\Gamma (\eta^{\prime} \to 2 \gamma) 
/ \Gamma (\eta \to 2 \gamma)$~\cite{Amsler:2008zz}
or of tensor meson decay widths~\cite{Amsler:1995td}.
The least square fit yields the fit parameters:
\eq 
\begin{array}{cccc} 
 \,\,\,  m_s  \,\,\, &  \,\,\,m_c \,\,\, & \,\,\, m_b  \,\,\, &   \\
\hline 
\,\,\,0.360\,\,\, & \,\,\,1.6\,\,\, &\,\,\, 4.8\,\,\, &  {\rm GeV} \\
\end{array}\;.
\en 
\eq\label{Lambdas}
\begin{array}{cccccccccccccccccc}
 \Lambda_{\rho/\omega/\phi} &  \Lambda_\eta & \Lambda_\eta^s & 
                               \Lambda_{\eta'} & \Lambda_{\eta'}^s &
 \Lambda_{K} &  \Lambda_{K^\ast} &   \Lambda_{D} &
 \Lambda_{D^\ast} &  \Lambda_{D_s} &   \Lambda_{D^\ast_s} &
 \Lambda_{B/B^\ast} &  \Lambda_{B_s/B_s^\ast} &  \Lambda_{J/\psi} &   
 \Lambda_{\eta_c/B_c} & \Lambda_{\Upsilon} &  \Lambda_{\eta_b} & \\
\hline
0.295 & 0.70 & 0.85 & 0.27 & 0.45 & 
0.87 & 0.30 & 1.4 & 2.3 & 1.95 & 2.6 &
 3.35 & 4.4 & 3.3 & 3.0 & 5.07 & 5.0 &  {\rm GeV}
\end{array}
\en 
\begin{table}[htpb]
\renewcommand{\arraystretch}{1.6}
     \setlength{\tabcolsep}{0.5cm}
     \centering
     \caption{Weak leptonic decay constants $f_{P(V)}$ in MeV.  
\label{tab:weak}}
\begin{tabular}{lcc}
\hline\hline 
Meson & Our & Data~\cite{Amsler:2008zz}\\ 
\hline\hline
$\pi^-$ & 130.2 & $130.4\pm 0.04 \pm 0.2$\\ 
\hline
$K^-$   & 155.4 & $155.5\pm 0.2 \pm 0.8$ \\ 
\hline
$D^+$   & 206.2 & $205.8 \pm 8.9$ \\ 
\hline
$D_s^+$ & 273.7 & $273\pm 10$ \\ 
\hline
$B^-$   & 216.4 & $216\pm 22$ \\ 
\hline
$B_s^0$ & 250.2 & $253 \pm 8 \pm 7$ \\ 
\hline
$B_c$   & 485.2 & $489\pm 5 \pm 3$~\cite{Chiu:2007bc}\\
\hline\hline
$\rho^+$       & 209.3 & $210.5 \pm 0.6$~\cite{Amsler:2008zz}\\
\hline
$D^{\ast +}$   & 187.0 & $245\pm 20^{+3}_{-2}$~\cite{Becirevic:1998ua}\\ 
\hline 
$D_s^{\ast +}$ & 213 & $272\pm 16^{+3}_{-20}$~\cite{Aubin:2005ar}\\ 
\hline
$B^{\ast +}$   & 210.6 & $196\pm24^{+39}_{-2}$~\cite{Becirevic:1998ua}\\ 
\hline
$B_s^{\ast 0}$ & 264.6 & $229\pm20^{+41}_{-16}$~\cite{Becirevic:1998ua}\\
\hline\hline
\end{tabular}
\end{table}
\begin{table}[htpb]
\renewcommand{\arraystretch}{1.6}
     \setlength{\tabcolsep}{0.5cm}
     \centering
     \caption{Electromagnetic leptonic decay constants $f_V$ 
of vector mesons with hidden  flavor in MeV.  
\label{tab:em}}
\begin{tabular}{lcc}
\hline
\hline 
Meson         & Our & Data~\cite{Amsler:2008zz} \\ \hline\hline
$\rho^0$      & 148.0 & 154.7 $\pm$ 0.7 \\ \hline
$\omega$      &  51.7 &  45.8 $\pm$ 0.8 \\ \hline
$\phi$        &  76.3 &  76   $\pm$ 1.2\\ \hline
$J/\psi$      & 277.4 & 277.6 $\pm$ 4\\ \hline
$\Upsilon(1s)$& 238.4 & 238.5 $\pm$ 5.5\\ \hline
\hline 
\end{tabular}
\end{table}
\begin{table}[htpb]
\renewcommand{\arraystretch}{1.4}
     \setlength{\tabcolsep}{0.5cm}
     \centering
     \caption{Electromagnetic and leptonic decay widths in keV. 
\label{tab:widths}}
\begin{tabular}{lcc}
\hline\hline 
Process & Our & Data~\cite{Amsler:2008zz}  \\\hline\hline
$\pi^0\to\gamma\gamma$&5.40 $\times$ 10$^{-3}$& 
$(7.7\pm0.6) \times  10^{-3}$\\ \hline
$\eta\to\gamma\gamma$              & 0.51 & $0.510\pm0.026$ \\\hline
$\eta^\prime\to\gamma\gamma$       & 4.27  & $4.28\pm0.19$ \\\hline
$\eta_c\to\gamma\gamma$            & 4.55 & $7.2\pm0.7\pm2.0$ \\\hline 
$\eta_b\to\gamma\gamma$            & 0.43 &\\ \hline
$\rho^0\to e^+e^-$                 & 6.33 &$7.04\pm0.06$ \\\hline
$\omega\to e^+e^-$                 & 0.76 &$0.60\pm0.02$ \\\hline
$\phi\to e^+e^-$                   & 1.27 &$1.27\pm0.04$ \\\hline
$J/\psi\to e^+e^-$                 & 5.54 &$5.55\pm0.14\pm0.02$ \\\hline
$\Upsilon\to e^+e^-$               & 1.34 &$1.34\pm0.018$ \\\hline
$\rho^{\pm}\to\pi^{\pm}\gamma$     & 72.42 &$68\pm7$ \\\hline
$\rho^0\to\eta\gamma$              & 63.25 &$62\pm17$ \\\hline
$\omega\to\pi^0\gamma$             & 682.71 &$703\pm25$ \\\hline
$\omega\to\eta\gamma$              & 7.63  &$6.1\pm2.5$ \\\hline
$\eta^\prime\to\omega\gamma$       & 12.44 &$9.06\pm2.87$ \\\hline
$\phi\to\eta\gamma$                & 51.72 &$58.9\pm0.5\pm2.4$ \\\hline
$\phi\to\eta^\prime\gamma$         & 0.41 &$0.27\pm0.01$ \\\hline
$K^{\ast \pm}\to K^\pm\gamma$      & 40.86 &$50\pm5$ \\\hline
$K^{\ast 0}\to K^0\gamma$          & 122.04 & $116\pm10$ \\\hline
$D^{\ast \pm}\to D^\pm\gamma$      & 0.62 & $1.5\pm0.8$ \\\hline
$D^{\ast 0}\to D^0\gamma$          & 20.27 & $<$ 0.9  $\times$ 10$^{3}$\\ 
\hline
$D^{\ast \pm}_s\to D^\pm_s\gamma$  & 0.30 & $<$ 1.8 $\times$ 10$^{3}$\\
\hline
$B^{\ast \pm}\to B^\pm\gamma$      & 0.36  &\\\hline
$B^{\ast 0}\to B^0\gamma$          & 0.12 &\\\hline 
$B^{\ast 0}_s\to B^0_s\gamma$      & 0.12 & \\\hline
$J/\psi \to \eta_c \gamma$         & 1.89 & 1.58 $\pm$ 0.37 \\\hline
$\Upsilon \to \eta_b \gamma$       & 0.02 & \\\hline
\hline
\end{tabular}
\end{table}
In Tables~\ref{tab:weak}--\ref{tab:widths} we list the results of our
numerical fit to the weak and electromagnetic leptonic decay constants 
together with their experimental values. 
Once the fit parameters are fixed one can use them to calculate 
a wide range of electromagnetic and dilepton decay widths. The results of
the calculation are presented in Table~\ref{tab:widths} which also includes
experimental results whenever they are available.

Note that the electromagnetic properties of the $\pi$, $K$, $\rho$ and $\omega$
mesons (form factors and radiative transitions) have also been considered
in the light-front constituent quark model (LFCQM) developed
in Ref.~\cite{Cardarelli:1995hn,Cardarelli:1995ap}. 
It was shown that a reasonable description of
data can be achieved with the use of the following values of the
constituent quark masses $m_u = m_d = 220$ MeV and $m_s = 410$ MeV,
which are close to our values $m_u = m_d = 217$ MeV and $m_s = 360$ MeV. 

\subsection{Dalitz decays}
 
In this section we apply our approach to the Dalitz decays 
$P \to \gamma l^+ l^-$ and $V \to P l^+ l^-$ (for a theoretical 
review, see e.g.~\cite{Landsberg:1986fd}). In particular, 
we analyze the transition form factors of the Dalitz decays 
$\eta\to\gamma\mu^+\mu^-$ and $\omega\to\pi^0\mu^+\mu^-$ decays.  
Both form factors have been measured 
the SERPUKHOV-134 Collaboration (Protvino)~\cite{Landsberg:1986fd} and 
recently by the NA60 experiment at the CERN SPS~\cite{Arnaldi:2009wb}. 
The $\omega\to\pi^0\mu^+\mu^-$ transition 
form factor has also been analyzed by the SND Collaboration at the 
BINP (Novosibirk)~\cite{Achasov:2008zz}.  

\begin{figure}[htbp]
\begin{center}
\hspace*{-0.5cm}
\begin{tabular}{c}
\includegraphics[scale=0.5]{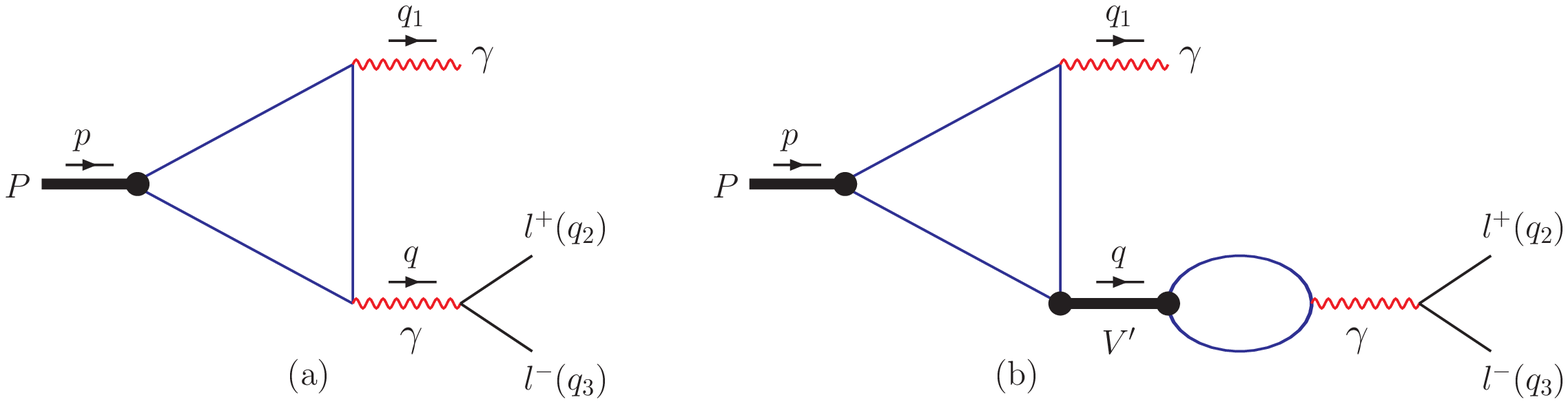} \\
\includegraphics[scale=0.5]{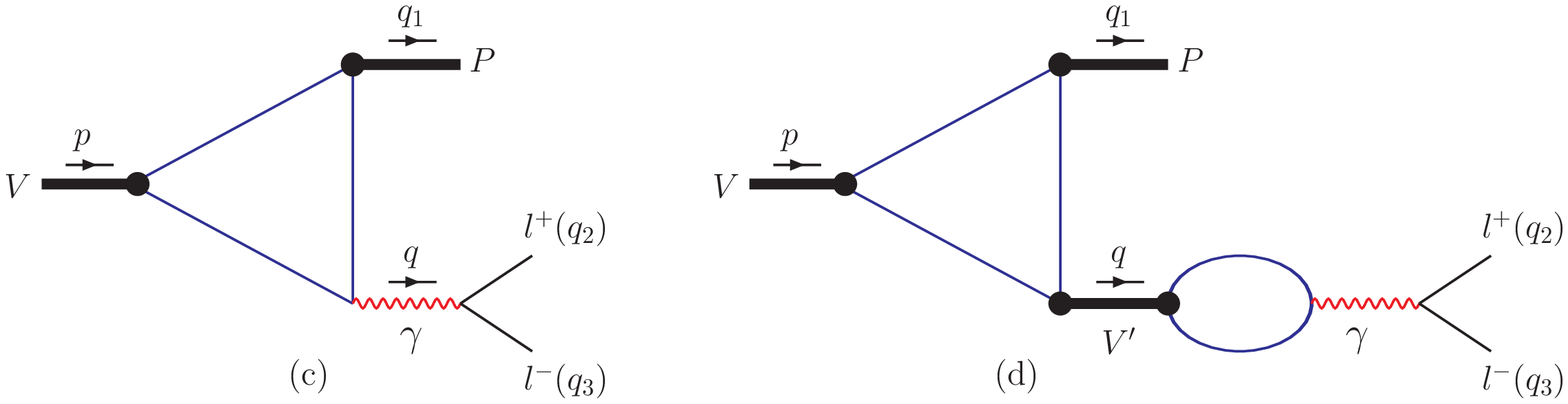}
\end{tabular}
\end{center}
\caption{
Diagrams describing the Dalitz decays 
$P \to \gamma l^+l^-$ (upper panel) 
and  $V \to Pl^+l^-$ (lower panel).
}
\label{fig:Dalitz}
\end{figure}

The diagrams describing the Dalitz decays are shown in Fig.~\ref{fig:Dalitz}.
They include both the diagrams with direct emission of  the photon
and resonance diagrams with an intermediate $V\to\gamma$ transition.
Note, that such decays have been studied in \cite{QCM}
by using a similar relativistic quark model with confinement 
(Quark Confinement Model). 
The differential cross sections w.r.t. the dilepton mass squared 
$q^2=(p_{l^+}+p_{l^-})^2$ reads
(the data is usually plotted w.r.t. the dilepton mass $M=\sqrt{q^2}$, see 
e.g. \cite{Arnaldi:2009wb})
\begin{eqnarray}
\frac{d\Gamma(P\to\gamma l^+l^-)}{dq^2} &=&
\frac23\frac{\alpha}{\pi}\frac{\Gamma(P\to\gamma\gamma)}{q^2}
\Big(1+\frac{2 m_l^2}{q^2}\Big)
\Big(1-\frac{4m_l^2}{q^2}\Big)^{1/2}
\Big(1-\frac{q^2}{m^2_P}\Big)^3
 \nonumber\\[2ex]
&\times& \Big|F_P(q^2)\Big|^2\,,\qquad 4m_l^2\le q^2\le m_P^2 \,,
\label{eq:PG}\\[5ex]
\frac{d\Gamma(V\to P l^+l^-)}{dq^2} &=&
\frac13\frac{\alpha}{\pi}\frac{\Gamma(V\to P\gamma)}{q^2}
\Big(1+\frac{2 m_l^2}{q^2}\Big) 
\Big(1-\frac{4m_l^2}{q^2}\Big)^{1/2}
\Big(1-\frac{q^2}{(m_V-m_P)^2}\Big)^{3/2}\,
\Big(1-\frac{q^2}{(m_V+m_P)^2}\Big)^{3/2}\, \nonumber\\[2ex]
&\times& \Big| F_{V}(q^2) \Big|^2\,, \qquad 4m_l^2\le q^2\le (m_V-m_P)^2 \,.
\label{eq:VP}
\end{eqnarray}
In our approach the normalized transition form factors are calculated 
in terms of the diagrams Fig.~\ref{fig:Dalitz}. The form factors
are given by
\begin{eqnarray}
F_P(q^2) &=& 
\frac{1}{g_{P\gamma\gamma}}\times
\Big\{
g_{P\gamma\gamma}(q^2) + 
\sum\limits_{V=\rho,\,\omega,\,\phi}\,
g_{PV\gamma}(q^2)\,D_V(q^2)\,q^2g_{V\gamma}(q^2)
\Big\}\,,
\label{eq:ffPg} \\[2ex]
F_{V}(q^2) &=& 
\frac{1}{g_{VP\gamma}}\times
\Big\{
g_{VP\gamma}(q^2) + 
\sum\limits_{V'=\rho,\,\omega,\,\phi}\,
g_{VPV'}(q^2)\,D_{V'}(q^2)\,q^2g_{V'\gamma}(q^2)
\Big\}\,.
\label{eq:ffVP} 
\end{eqnarray}
The decay constants $g_{P\gamma\gamma}\equiv g_{P\gamma\gamma}(0)$ etc.,
are the same as defined in Sec.~\ref{sec:fit}. One has to note that
if we assume that the quantities $g(q^2)$ in Eqs.~(\ref{eq:ffPg}) and 
(\ref{eq:ffVP}) do not depend on $q^2$ then we reproduce the results of
vector meson dominance (VMD).
  
\begin{figure}[htbp]
\begin{center}
\hspace*{-0.5cm}
\begin{tabular}{lcr}
\includegraphics[scale=0.48]{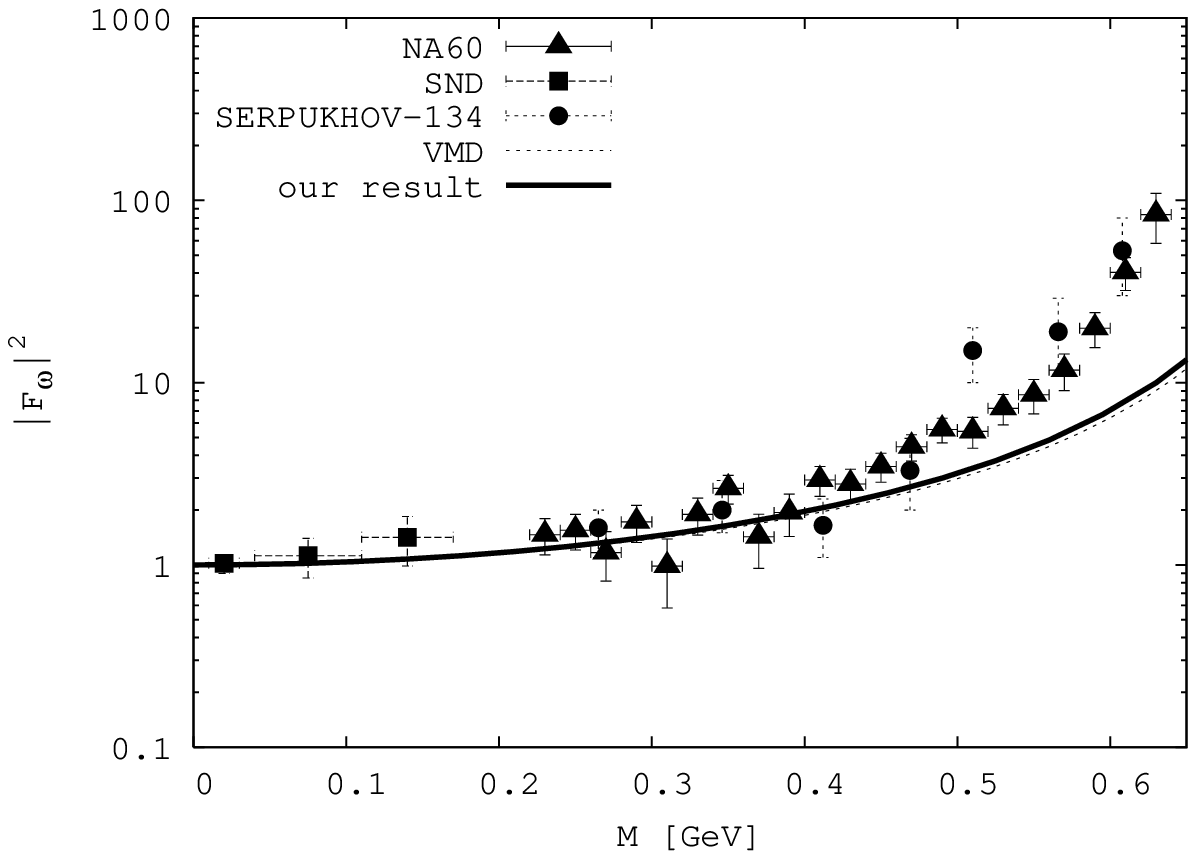} &
\includegraphics[scale=0.48]{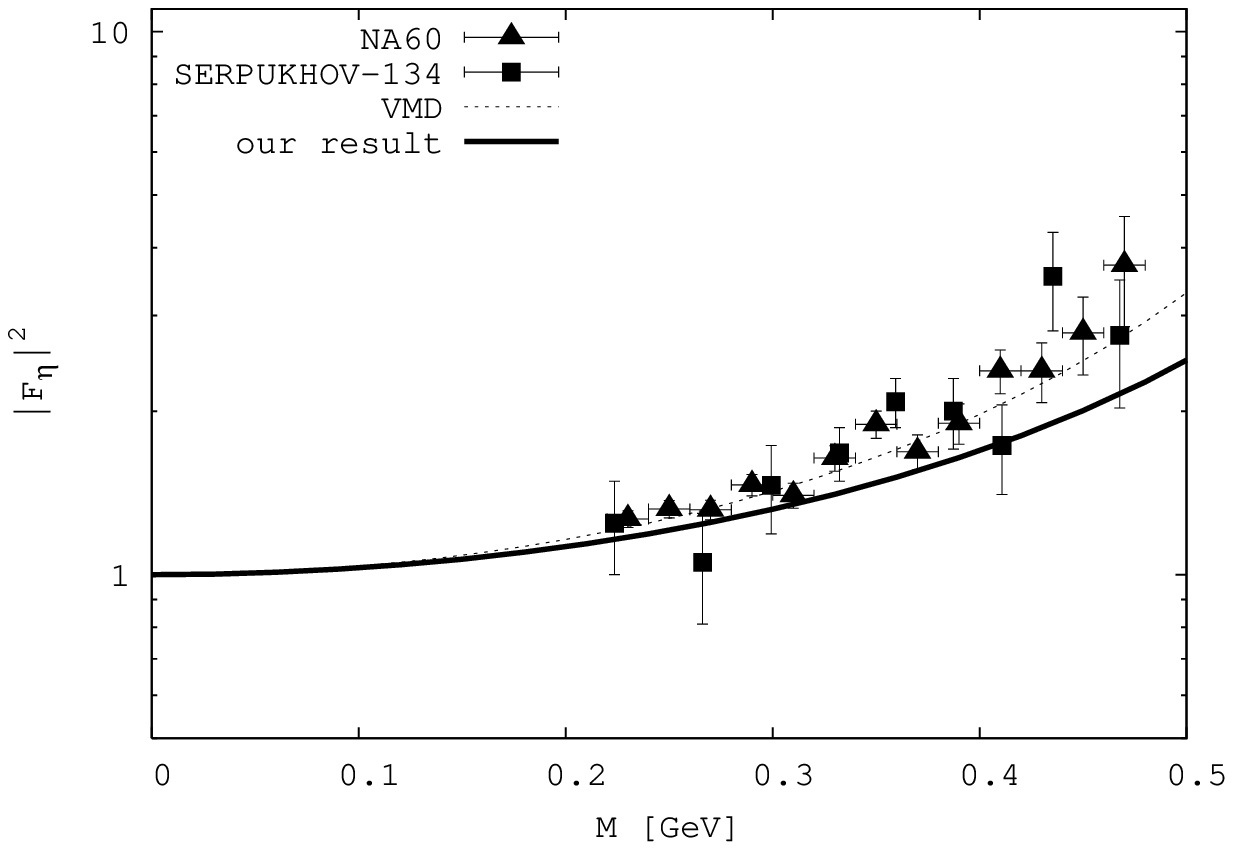} &
\includegraphics[scale=0.48]{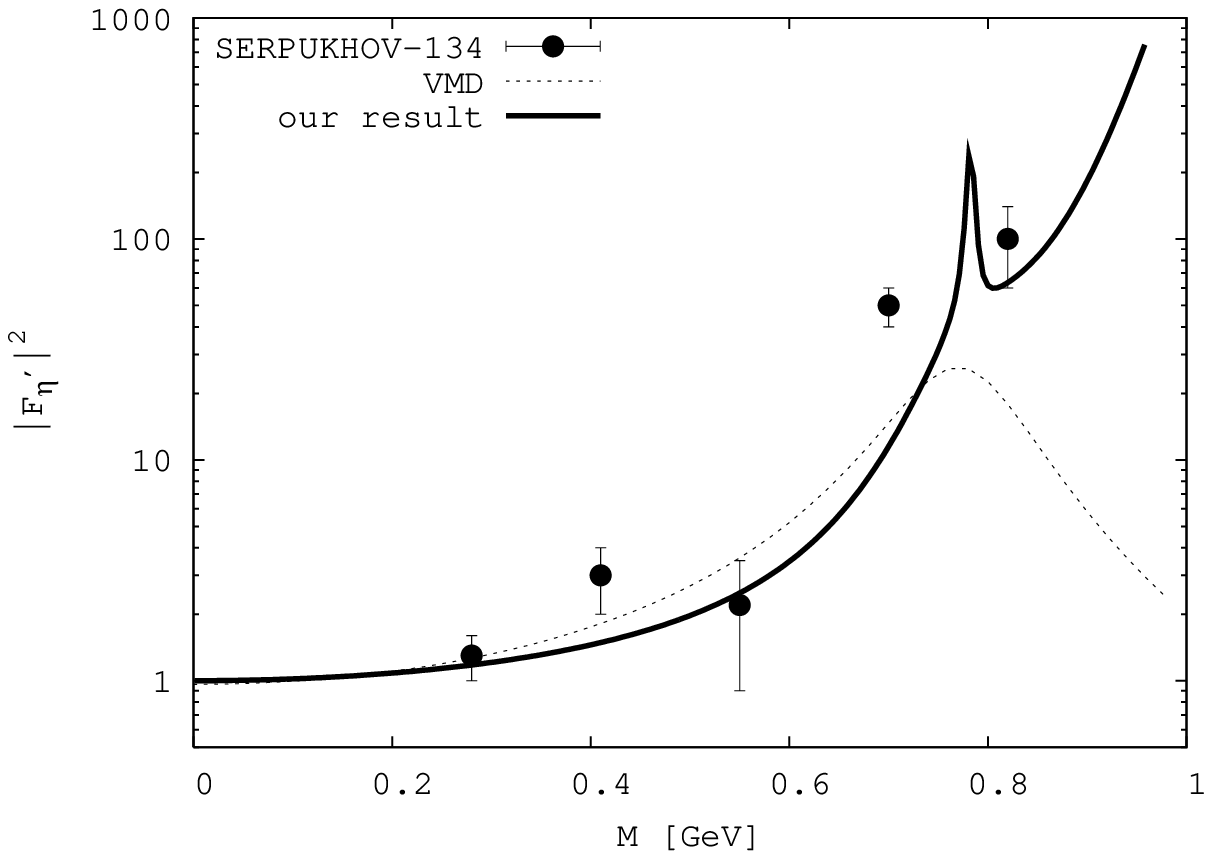} 
\end{tabular}
\end{center}
\caption{
The calculated form factors of the Dalitz decays 
$\omega\to\pi^0 l^+l^-$, $\eta\to\gamma l^+l^-$ and $\eta'\to\gamma l^+l^-$
as functions of the dilepton mass $M=\sqrt{q^2}$.
Experimental data are taken from \cite{Landsberg:1986fd,Arnaldi:2009wb} 
for $\omega\to\pi^0 \mu^+\mu^-$ and  $\eta\to\gamma \mu^+\mu^-$, 
from \cite{Achasov:2008zz} for $\eta\to\gamma e^+e^-$ and
from \cite{Landsberg:1986fd} for $\eta'\to\gamma e^+e^-$ 
(the five uncorrected by background points are included).  
For comparison we plot the VMD-curves.}
\label{fig:ff}
\end{figure}

The contributions of the intermediate vector meson resonances are described 
by a Breit--Wigner form where we have used a constant width in the imaginary 
part, i.e. we have used
\begin{equation}
D_V(q^2) = \frac{1}{m^2_V-q^2-i m_V\Gamma_V} \,.
\end{equation}  
In Fig.~\ref{fig:ff}  we plot the transition
form factors for the two Dalitz decays
$\eta\to\mu^+\mu^-\gamma$ and $\omega\to\pi^0 \mu^+\mu^-$ 
measured by  
the SERPUKHOV-134 Collaboration (Protvino)~\cite{Landsberg:1986fd}, 
the CERN NA60 Collaboration~\cite{Arnaldi:2009wb} and 
the SND Collaboration  BINP (Novosibirk)~\cite{Achasov:2008zz}. 
For completeness we also display our prediction for the Dalitz decay 
$\eta' \to \gamma e^+ e^-$ in comparison to the older data given 
in~\cite{Landsberg:1986fd}.   
In Table~\ref{tab:slope}  we present our results for 
the slope parameters defined by $F'_X(0)$.  
Finally, our predictions for the Dalitz decay widths are given
in Table~\ref{tab:width}. The agreement with the data is generally
quite good except for the decay $\pi^0 \to \gamma e^+e^-$ which comes out
too small in our model. This is not unexpected since the decay
$\pi^0 \to \gamma \gamma$ also comes out too small in our model (see Table IV).

\begin{table}[htpb]
\renewcommand{\arraystretch}{1.3}
     \setlength{\tabcolsep}{0.5cm}
     \centering
     \caption{Slope parameters of the Dalitz transition form factors
in GeV$^{-2}$.  }
\label{tab:slope}
\begin{tabular}{lcc}
\hline
\hline
Decay mode    & Our   & Data  \\ \hline\hline
$\pi^0 \to \gamma l^+ l^-$  & 1.4 & $5.5 \pm 1.6$~\cite{Fischer:1977zd} \\
\hline
$\eta \to \gamma l^+ l^-$   & 1.4 & $3 \pm 1$~\cite{Bushnin:1978fq} \\
                            &     &
                     $1.95 \pm 0.17 \pm 0.05$~\cite{Arnaldi:2009wb}\\
\hline
$\eta^\prime \to \gamma l^+ l^-$  & 1.1 & 1.68~\cite{Dzhelyadin:1979za} \\
\hline
$\rho^0 \to \pi^0 l^+ l^-$            & 1.9 & \\
\hline
$\omega \to \pi^0 l^+ l^-$            & 1.9 &
  $2.24 \pm 0.06 \pm 0.02$~\cite{Arnaldi:2009wb}\\
\hline
$\rho^0 \to \eta l^+ l^-$               & 2.2 & \\
\hline
$\omega \to \eta l^+ l^-$             & 2.2 & \\
\hline
$\phi \to \eta l^+ l^-$               & 1.4 & \\
\hline
$\phi \to \eta' l^+ l^-$              & 2.5 & \\
\hline\hline
\end{tabular}
\end{table}
\begin{table}[htpb]
\renewcommand{\arraystretch}{1.3}
     \setlength{\tabcolsep}{0.5cm}
     \centering
     \caption{Dalitz decay widths in keV.}
\label{tab:width}
\begin{tabular}{llc}
\hline
\hline
Decay mode    & Our   & Data~\cite{Amsler:2008zz}  \\ \hline\hline
$\pi^0 \to \gamma e^+ e^-$         & 6.4 $\times$ $10^{-5}$ 
                                   & $(9.39 \pm 0.72)$ $\times$ $10^{-5}$\\
\hline
$\eta        \to \gamma e^+ e^-$   & 8.5  $\times$ $10^{-3}$  
                                   & $(8.84 \pm 1.14)$ $\times$ $10^{-3}$
\\
\hline
$\eta        \to \gamma \mu^+ \mu^-$  & 0.4 $\times$ $10^{-3}$
                                      & $(0.40 \pm 0.06)$ $\times$ $10^{-3}$ \\
\hline
$\eta^\prime \to \gamma e^+ e^-$    & 9.0 $\times$ $10^{-2}$ & $ < 0.2 $ \\
\hline
$\eta^\prime \to \gamma \mu^+ \mu^-$& 2.0 $\times$ $10^{-2}$ 
                                    & $(3.2 \pm 1.2)$ $\times$ $10^{-2}$  \\
\hline
$\rho^0 \to \pi^0 e^+ e^-$         & 0.66 & $ < 2.3 $ \\
\hline
$\rho^0 \to \pi^0 \mu^+ \mu^-$     & 7.0 $\times$ $10^{-2}$ & \\
\hline
$\omega \to \pi^0 e^+ e^-$         & 6.23 & $6.54 \pm 0.77$ \\
\hline
$\omega \to \pi^0 \mu^+ \mu^-$     & 0.65 & $0.82 \pm 0.20$ \\
\hline
$\rho^0 \to \eta e^+ e^-$          & 0.46 & $ < 1$ \\
\hline
$\omega \to \eta e^+ e^-$          & 5.6 $\times$ $10^{-2}$  
                                   & $ < 9.3$ $\times$ $10^{-2}$ \\
\hline
$\phi   \to \eta e^+ e^-$          & 0.44 & $0.49 \pm 0.04$ \\
\hline
$\phi \to \eta' e^+ e^-$           & 2.1 $\times$ $10^{-3}$    &  \\
\hline
$\phi \to \eta \mu^+ \mu^-$        & 2.2 $\times$ $10^{-2}$
                                   & $ < 4 $ $\times$ $10^{-2}$ \\
\hline\hline
\end{tabular}
\end{table}

\section{Summary} 
\label{sec:summary}

We have refined a relativistic constituent quark model developed
in our previous papers to include quark confinement effects.
Quark confinement was implemented by introducing an upper cutoff on a scale 
integration which, in the original quark model, extends to infinity. The
introduction of such an infrared cutoff removes all physical quark 
thresholds in the original quark diagrams. 
The cutoff parameter is taken to be the same for all physical processes.
We adjust the model parameters by fitting the calculated
quantities of the basic physical processes to available experimental data.
As an application, we calculate the electromagnetic form factors of pion
and the transition form factors of the Dalitz decays  
$P\to\gamma l^+l^-$ and $V\to P l^+l^-$.
We extend our approach to mesons containing strange, charm and bottom quarks
and calculate their leptonic and radiative decay constants.

\begin{acknowledgments}

We very much appreciate the help of Sanja Damjanovic and Hans Specht  
who provided us with the experimental data on the Dalitz decays. 
This work was supported by the DFG under Contract
No. FA67/31-2 and No. GRK683. M.A.I. and J.G.K. appreciate the partial
support of the Heisenberg-Landau program. The work of M.A.I. was supported by
the DFG grant KO 1069/12-1. M.A.I. appreciates the support of
Forschungszentrum "Elementarkr\"afte und mathematische Grundlagen".
This research is also part of the European Community-Research
Infrastructure Integrating Activity ``Study of Strongly Interacting
Matter'' (HadronPhysics2, Grant Agreement No. 227431) and of 
the President grant of Russia ``Scientific Schools'' No. 3400.2010.2.   
The work is partially supported by Russian Science and Innovations 
Federal Agency under contract  No. 02.740.11.0238. 

\end{acknowledgments}

\appendix\section{Loop integration techniques}

In order to demonstrate how the loop integrations are done 
consider a $n$-point one-loop diagram with $n$ local propagators 
$S_i(k + v_i)$ and $n$ Gaussian vertex functions
$\Phi_{i} \left( -(k+v_{i+n})^2\right)$. In Minkowski space the one-loop
diagram can be written as
\begin{equation}
\label{eq:diag_loop} 
I_n(p_1,...,p_n) = \int\!\! \frac{d^4k}{\pi^2i}\,  {\rm tr} \,
\prod\limits_{i=1}^n \Phi_{i} \left( -(k+v_{i+n})^2\right)\,
\Gamma_i\, S_i(k+v_i)
\end{equation}
where the vectors $v_i$ are  linear combinations of the external
momenta $p_i$ to be specified in the following, $k$ is the loop momentum,
and the $\Gamma_i$ are Dirac matrices (or strings of Dirac matrices) for the 
$i$th meson. 
The external momenta $p_i$ are all chosen to be ingoing such that one has 
$\sum\limits_{i=1}^n p_i=0$. Due to translational invariance 
the integral Eq.~(\ref{eq:diag_loop}) is invariant under a
shift of the loop momentum $k\to k+ l$ by any four-vector $l$.
The four-vector $l$ may be any linear combination of the external
momenta $p_i$.  
 
Using the Schwinger representation of the local quark propagator one has
\begin{eqnarray}
S_i(k+v_i) &=& (m_i+\not\! k+\not\!v_i)
\times\int\limits_0^\infty\! 
d\beta_i\,\exp\left[-\beta_i\,(m^2_i-(k+v_i)^2)\right]\,.
\end{eqnarray}
For the vertex functions one takes the Gaussian form. One has
\begin{eqnarray}
\label{eq:vert} 
\Phi_{i} \left( - (k+v_{i+n})^2\right) &=&
\exp\left[\beta_{i+n}\,(k+v_{i+n})^2\right] \qquad i=1,...,n\, ,
\end{eqnarray}
where the parameters $\beta_{i+n}=s_{i}=1/\Lambda^2_{i}$ are 
related to the size parameters. 
The numerator factors
$m_i\, + \!\!\not\! k\, + \!\!\not\! v_i$ 
 can be replaced by a
differential operator in the following manner:

\begin{eqnarray}
\label{eq:dif} 
I_n(p_1,...,p_n) &=& 
\int\!\! \frac{d^4k}{\pi^2i}\,{\rm tr} \,  
 \prod\limits_{i=1}^n\int\limits_0^\infty d\beta_i\, e^{-\beta_i m^2_i}
(m_i+\not\! k+\not\!v_i)\times
\exp\left\{\sum\limits_{i=1}^{2\,n}\beta_i\,(k+v_i)^2\right\} 
\nonumber\\[3ex]
&=&
\int\!\! \frac{d^4k}{\pi^2i}\,{\rm tr} \,  
 \prod\limits_{i=1}^n\int\limits_0^\infty d\beta_i\, e^{-\beta_i m^2_i}
\left(m_i+\not\!v_i+\frac{1}{2}\not\!\partial_r\right)\times
\exp\left\{\beta k^2+2kr+\sum\limits_{i=1}^{2\,n}\beta_i\,v_i^2\right\}
\end{eqnarray}
where  $\beta=\sum\limits_{i=1}^{2\,n}\beta_i$ and 
$r=\sum\limits_{i=1}^{2\,n}\beta_i\,v_i$. Next one does the loop integration
and moves the Gaussian $\exp\Big(-r^2/\beta\Big)$ to the left  
of the differential operator. By using the identity
\begin{equation}
\label{eq:identity}
-\frac{r^2}{\beta}+\sum\limits_{i=1}^{2\,n}\beta_i\,v_i^2
=\frac{1}{\beta} \sum\limits_{1\le i<j
\le 2\,n}\beta_i\,\beta_j\,(v_i-v_j)^2
\end{equation}
one obtains
\begin{eqnarray}
I_n(p_1,...,p_n) &=& \prod\limits_{i=1}^n
\int\limits_0^\infty\frac{d\beta_i}{\beta^2} \times
\exp\left\{-\sum\limits_{i=1}^{n}\beta_i m^2_i
+\frac{1}{\beta}
\sum\limits_{1\le i<j\le 2\,n}\beta_i\,\beta_j\,(v_i-v_j)^2\right\}
\nonumber\\[2ex]
&\times& {\rm tr}\, \prod\limits_{i=1}^n \Gamma_i
\left(m_i+\not\! v_i-\frac{1}{\beta}\not\! r
+\frac{1}{2}\not\!\partial_r\right)\,.
\end{eqnarray}
We have written a FORM \cite{Vermaseren:2000nd} program that achieves 
the necessary commutations in a very efficient way.
As described in the main text the set of Schwinger parameters 
$\beta_i\,\, (i=1,...,n)$
can be  turned into a simplex by writing
\begin{equation}
\label{eq:simplex}
\int\limits_0^\infty\!
d^n\beta F(\beta_1,...,\beta_n) = \int\limits_0^\infty\! dt\,t^{n-1}
\int\! d^n\alpha\,
\delta\left(1-\sum\limits_{i=1}^n \alpha_i\right)
F(t\,\alpha_1,...,t\,\alpha_n)\,.
\end{equation} 
One then arrives at the final representation of the one-loop n-point
diagram in the form
\begin{eqnarray}
I_n(p_1,...,p_n) &=& \int\limits_0^\infty\! dt
\frac{t^{n-1}}{(s+t)^2} \int\! d^n\alpha\,
\delta\Big(1-\sum\limits_{i=1}^n \alpha_i\Big)
\exp\Big\{ - t\,z_{\,\rm loc} + \frac{s\,t}{s+t}\,z_1 + \frac{s^2}{s+t}\,z_2 \Big\}
\nonumber\\[2ex]
&\times& {\rm tr}\, \prod\limits_{i=1}^n
\Gamma_i \left(m_i+\not\! v_i-\frac{1}{s+t}\not\! r
+\frac{1}{2}\not\!\partial_{r}\right)\, ,
\label{eq:master}
\end{eqnarray}
where
\begin{eqnarray}
\label{eq:z-form} 
z_{\,\rm loc} &=&
\sum_{i=1}^n\alpha_i\,m^2_i -\sum_{1\le i<j\le n}\alpha_i\,\alpha_j\,A_{ij}\, ,
\nonumber\\[2ex]
z_1&=&
\sum_{i=1}^n\alpha_i\,\sum_{j=n+1}^{2\,n}\bar\beta_j\,A_{ij}
-\sum_{1\le i<j\le n}\alpha_i\,\alpha_j\,A_{ij}\, ,
\nonumber\\[2ex]
z_2 &=&
\sum_{n+1 \le i<j\le 2\,n}\bar\beta_i\,\bar\beta_j\,A_{ij}\,, 
\nonumber\\[4ex]
r &=& t\,\sum_{i=1}^n\alpha_i\,v_i
  +s\,\sum_{i=n+1}^{2\,n}\bar\beta_i\,v_i\,.
\nonumber
\end{eqnarray}
Here, $\bar\beta_{i+n} = s_i/s,\,\, s=\sum\limits_{i=1}^n s_i$.
The matrix $A_{ij}=(v_i-v_j)^2 \quad (1\le i,j \le 2n)$  depends 
on the invariant variables of the process.

There are altogether $n$ numerical integrations, $n-1$ $\alpha$--parameter
integrations and the integration over the scale parameter $t$. For the 
derivative of the
(two-point function) mass operator one has to do one more $\alpha$--parameter
integration due to the extra propagator which comes in after the
differentiation. The integration of the derivative 
of the mass operator proceeds in analogy to the $n$ point function case
described in this Appendix. We mention that the correctness of the numerical 
integration procedure can be checked very conveniently by shifting the
loop integration momentum by a fixed momentum four-vector. The numerical
evaluations have been done by a numerical program written in the FORTRAN code.

Some further remarks are in order. The
convergence of the loop integral Eq.~(\ref{eq:master}) is defined by the
local $\alpha$ form $z_{\,\rm loc}$. If  $z_{\,\rm loc}\le 0$ the
$t$-integration becomes divergent due to contributions from the large 
$t$-region. 
The large $t$-region corresponds to that region where the singularities
of the diagram with its local
quark propagators appear. However, as described before, if one introduces 
an infrared cutoff on the upper limit of the t-integration, all 
singularities vanish because the integral is now convergent for any value
of the set of kinematical variables. Note that the one-loop integration 
techniques described in this appendix can be extended to an arbitrary 
number of loops in a straightforward manner. Of particular interest is the
extension to the two-loop case needed for the description of baryon 
transitions.

\section{Gauge invariance of the \boldmath{$\rho^{0}\to\gamma$} transition}

In this appendix we want to demonstrate by an explicit calculation that the 
transition amplitude $\rho^0\to\gamma$ written down in Eq.~(\ref{VGint})  
satisfies gauge invariance, i.e. that e.g. the on-shell photon has only two 
transverse degrees of freedom. According to the formalism developed in
Sec.~\ref{sec:framework}~B there are the two contributions to the transition 
$\rho^{0}\to\gamma$.
The first contribution results from the minimal substitution in the free
quark Lagrangian and is depicted in Fig.~\ref{fig:MW}(a). In
Fig.~\ref{fig:MW}(b) we depict the point interaction
contribution resulting from gauging the nonlocal Lagrangian.

We begin by considering the transition of an on-shell $\rho$ to an
off-shell photon with invariant mass $p^2$. The corresponding transition
amplitude $M^{\mu\nu}(p)$ must satisfy the gauge invariance condition
$p_\nu M^{\mu\nu}(p)=0$ as has already been assumed in writing down 
Eq.~(\ref{VGint}). We shall now show by explicit calculation that the
non-gauge invariant pieces in the two contributions cancel each other
resulting in an overall gauge invariant contribution. First we isolate
the non-gauge invariant pieces in the two respective contributions by 
writing
\begin{eqnarray}
M_{a}^{\mu\nu}(p) &=&
\int\frac{d^4k}{4\pi^2i}\Phi_\rho(-k^2)
\,\text{tr}\Big(\gamma^\mu S(k+\tfrac12\, p)\gamma^\nu S(k-\tfrac12\, p)\Big)
\nonumber\\[2ex] 
&=& g^{\mu\nu}\Big[I_{a}^{(1)}(p^2)+ I_{a}^{(2)}(p^2)\Big]
+(g^{\mu\nu}p^2 - p^\mu p^\nu) \,I_{a}^\perp(p^2)\,.
\end{eqnarray}
The non-gauge invariant contributions are given by
\begin{eqnarray}
I_{a}^{(1)}(p^2) &=& 
\int\limits_0^\infty \frac{dt}{(s+t)^2} e^{-z_1} \,, \qquad
z_1 = tm^2 - \frac{s\,t}{s+t} \frac{p^2}{4}\,, 
\nonumber\\[2ex] 
I_{a}^{(2)}(p^2) &=&
\int\limits_0^\infty \frac{dt\, t}{(s+t)^2} \int\limits_0^1\! 
d\alpha e^{-z_2} \biggl[ - \frac{1}{s+t} 
+ \frac{2\,t^2}{(s+t)^2} \biggl(\alpha - \frac{1}{2}\biggr)^2 p^2 \biggr]\,, 
\nonumber\\[2ex]
z_2&=& t\, \Big(m^2 - \alpha(1-\alpha)p^2\Big) 
- \frac{s\,t}{s+t} \biggl(\alpha - \frac{1}{2}\biggr)^2 p^2\,, 
\end{eqnarray}
whereas the gauge invariant contribution is given by
\begin{equation} 
I_{a}^\perp(p^2) =
\int\limits_0^\infty \frac{dt\, t}{(s+t)^2} \int\limits_0^1\! 
d\alpha e^{-z_2} \biggl[\frac12
-\frac{2\,t^2}{(s+t)^2} \biggl(\alpha - \frac{1}{2}\biggr)^2 \biggr]\,.
\end{equation}
Similarly we have
\begin{eqnarray} 
M_{b}^{\mu\nu}(p)&=& -\,\int\frac{d^4k}{4\pi^2i} \, 
(2k+\tfrac12\, p)^\mu \,
\int\limits_0^1 d\alpha \, 
\Phi'_\pi\Big(-\alpha\,(k+\tfrac12\, p)^2-(1-\alpha)\,k^2\Big)
\text{tr}\Big(\gamma^\nu S(k)\Big) 
\nonumber\\[2ex] 
&=& g^{\mu\nu}\, I_{b}^{(3)}(p^2)
+ (g^{\mu\nu}p^2 - p^\mu p^\nu) \,I_{b}^\perp(p^2)\,,
\nonumber\\[2ex] 
 I_{b}^{(3)}(p^2) &=& 
\int\limits_0^\infty \frac{dt\,s}{(s+t)^2} \int\limits_0^1\! 
d\alpha\, e^{-z_3} \biggl[ - \frac{1}{s+t} 
- \biggl(1 - \frac{2\,s\,\alpha}{s+t}\biggr)\frac{s\,\alpha}{s+t}  
\frac{p^2}{4}\biggr]\,, 
\nonumber\\[2ex] 
 I_{b}^\perp(p^2) &=& 
\int\limits_0^\infty \frac{dt\,s}{(s+t)^2}  \int\limits_0^1\! 
d\alpha\, e^{-z_3} \frac14 \biggl(1-  \frac{2\,s\,\alpha}{s+t}\biggr)
               \frac{s\,\alpha}{s+t}\,,\qquad 
z_3 = tm^2 - \Big(1-\frac{s\,\alpha}{s+t}\Big)\frac{s\,\alpha}{4}p^2\,. 
\end{eqnarray} 
Here $s=1/\Lambda^2_\rho$. The non-gauge invariant pieces
$I_{a}^{(1,2)}$ and $I_{b}^{(3)}$
cancel each other as can be seen by the following transformations.
First, we note that the integrands of the integrals
$I_{a}^{(2)}(p^2)$ and $I_{b}^{(3)}(p^2)$
may be expressed via the derivatives of $z_2$ and $z_3$, respectively.
\eq
I_{a}^{(2)}(p^2)  &\longrightarrow&
 - \frac{1}{s+t} 
+ \frac{2\,t^2}{(s+t)^2} \biggl(\alpha - \frac{1}{2}\biggr)^2 p^2 
= - \frac{1}{s+t} 
\biggl[ 1 - \biggl(\alpha - \frac{1}{2} \biggr)  
\frac{\partial  z_2}{\partial\alpha} \biggr] \,, 
\nonumber\\[2ex]
I_{b}^{(3)}(p^2) &\longrightarrow&
-\frac{1}{s+t} - \frac{p^2}{4} \biggl( 1 - \frac{2\,\alpha\,s}{s+t} \biggr) 
\frac{\alpha\,s}{s+t} = - \frac{1}{s+t} 
\biggl[ 1 - \alpha \frac{\partial  z_3}{\partial\alpha} \biggr]\,. 
\nonumber
\en 
The $\alpha$-integration can be done by using 
the bound-state conditions
$z_2(\alpha=1) = z_2(\alpha=0) = z_3(\alpha=1)=z_1$.
One obtains
\eq 
I_{a}^{(2)}(p^2) &=& 
- \int\limits_0^\infty \frac{dt\, t}{(s+t)^3} \int\limits_0^1 
d\biggl[\biggl(\alpha - \frac{1}{2}\biggr) \, e^{- z_2}\biggr] = 
 - \int\limits_0^\infty \frac{dt t}{(s+t)^3} e^{- z_1}\,, 
\nonumber\\[2ex]
I_{b}^{(3)}(p^2) &=& - \int\limits_0^\infty \frac{dt\,s}{(s+t)^3} 
\int\limits_0^1 d\biggl[\alpha \, e^{- z_3}\biggr] = 
- \int\limits_0^\infty \frac{dt\,}{(s+t)^3} e^{- z_1} 
\nonumber
\en 
Finally, one has
\eq 
I_{a}^{(1)}(p^2)+I_{a}^{(2)}(p^2)+I_{b}^{(3)}(p^2)=
\int\limits_0^\infty \frac{dt}{(s+t)^2} e^{- z_1} 
\biggl[ 1 - \frac{t}{s+t} - \frac{s}{s+t} \biggr] \equiv 0\,. 
\en 
The gauge invariance condition $p_\nu M^{\mu\nu}(p)=0$ also guarantees
that the longitudinal component of the photon decouples as $p^2 \to 0$.

\end{document}